\documentclass[10pt,aps,prb,twocolumn]{revtex4-1}   % style for Physical Review B and AJP are similar %\documentclass[10pt,a4paper]{article}   % style for Physical Review B and AJP are similar %\documentclass[12pt,aps,prb,preprint]{revtex4}   % style for Physical Review B and AJP are similar

\usepackage{amsmath}    % need for subequations
\usepackage{amsfonts}  %note how statements can be commented out
\usepackage{amssymb}
\usepackage{graphicx}   % for figures

\usepackage[latin1]{inputenc}     % pour les accents
\usepackage[T1]{fontenc}

\usepackage{float}   % pour que les images soient inclusent dans le texte

\graphicspath{{./images/}}

\begin{document}

\title{Measurement in the de Broglie-Bohm interpretation: \\Double-slit, Stern-Gerlach and EPR-B}
\author{Michel Gondran}
 \affiliation{University
Paris Dauphine, Lamsade, 75 016 Paris, France}
 \email{michel.gondran@polytechnique.org}   %optional

 \author{Alexandre Gondran}
 \affiliation{\'Ecole Nationale de l'Aviation Civile, 31000 Toulouse, France}
  \email{alexandre.gondran@enac.fr}

\begin{abstract}

We propose a pedagogical presentation of measurement in the de Broglie-Bohm interpretation. In this heterodox interpretation, the position of a quantum particle exists and is piloted by the phase of the wave function. We show how this position explains  determinism and realism in the three most important experiments of quantum measurement: double-slit, Stern-Gerlach and EPR-B. 

First, we demonstrate the conditions in which the de Broglie-Bohm interpretation can be assumed to be valid through continuity with classical mechanics. 

Second, we present a numerical simulation of the double-slit experiment performed by Jönsson in 1961 with electrons. It demonstrates the continuity between classical mechanics and quantum mechanics: evolution of the probability density  at various distances and convergence of the quantum trajectories to the classical trajectories when h tends to 0. 

Third, we present an analytic expression of the wave function in the Stern-Gerlach experiment. This explicit solution requires the calculation of a Pauli spinor with a spatial extension. This solution enables to demonstrate the decoherence of the wave function and the three postulates of quantum measurement: quantization, the Born interpretation and  wave function reduction. The spinor spatial extension also enables the introduction of the de Broglie-Bohm trajectories, which gives a very simple explanation of the particles' impact and of the measurement process. 

Finally, we study the EPR-B experiment, the Bohm version of the Einstein-Podolsky-Rosen experiment. Its theoretical resolution in space and time shows that a causal interpretation exists where each atom has a position and a spin. This interpretation avoids the flaw of the previous causal interpretation. We recall that a physical explanation of non-local influences is possible.

\end{abstract}

\maketitle

%%%%%%%%%%%%%%%%%%%%%%%%%%%%%%%%%%%%%%%%%%%%%%%%%%%%%%%%%%%%%%%%%%%%%%%%%%%%%%
%%%%%%%%%%%%%%%%%%%%%%%%%%%%%%%%%%%%%%%%%%%%%%%%%%%%%%%%%%%%%%%%%%%%%%%%%%%%%%
%%%%%%%%%%%%                    INTRODUCTION                     %%%%%%%%%%%%%
%%%%%%%%%%%%%%%%%%%%%%%%%%%%%%%%%%%%%%%%%%%%%%%%%%%%%%%%%%%%%%%%%%%%%%%%%%%%%%
%%%%%%%%%%%%%%%%%%%%%%%%%%%%%%%%%%%%%%%%%%%%%%%%%%%%%%%%%%%%%%%%%%%%%%%%%%%%%%
\section{Introduction}

"\emph{I saw the impossible done}".\cite{Bell_1982} This is how John Bell describes his
inexpressible surprise in 1952 upon the publication of an article
by David Bohm~\cite{Bohm_1952}. The impossibility came from a
theorem by John von Neumann outlined in 1932 in his book \emph{The
Mathematical Foundations of Quantum Mechanics},\cite{vonNeumann}
which seemed to show the impossibility of adding "hidden
variables" to quantum mechanics. This impossibility, with its
physical interpretation, became almost a postulate of quantum
mechanics, based on von Neumann's indisputable authority as a
mathematician. As Bernard d'Espagnat notes in 1979:

"\emph{At the university, Bell had, like all of us, received from
his teachers a message which, later still, Feynman would
brilliantly state as follows: "No one can explain more than we
have explained here [...]. We don't have the slightest idea of a
more fundamental mechanism from which the former results (the
interference fringes) could follow". If indeed we are to believe
Feynman (and Banesh Hoffman, and many others, who expressed the
same idea in many books, both popular and scholarly), Bohm's
theory cannot exist. Yet it does exist, and is even older than
Bohm's papers themselves. In fact, the basic idea behind it was
formulated in 1927 by Louis de Broglie in a model he called "pilot
wave theory". Since this theory provides explanations of what, in
"high circles", is declared inexplicable, it is worth
consideration, even by physicists [...] who do not think it gives us the final answer to the
question "how reality really is.}"\cite{dEspagnat}

And in 1987, Bell wonders about his teachers' silence concerning
the Broglie-Bohm pilot-wave:

"\emph{But why then had Born not told me of this 'pilot wave'? If
only to point out what was wrong with it? Why did von Neumann not
consider it? More extraordinarily, why did people go on producing
"impossibility" proofs after 1952, and as recently as 1978?
While even Pauli, Rosenfeld, and Heisenberg could produce no more
devastating criticism of Bohm's version than to brand it as
"metaphysical" and "ideological"? Why is the pilot-wave picture
ignored in text books? Should it not be taught, not as the only
way, but as an antidote to the prevailing complacency? To show
that vagueness, subjectivity and indeterminism are not forced on
us by experimental facts, but through a deliberate theoretical
choice?}"\cite{Bell_1987}

More than thirty years after John Bell's questions, the
interpretation of the de Broglie-Bohm pilot wave is still ignored by
both the international community and the textbooks.

What is this pilot wave theory? 
For de Broglie, a quantum particle is not only defined by its wave function.
He assumes that the quantum particle also has a position which is piloted by the wave function.\cite{Broglie_1927} 
However only the probability density of this position is known.
The position exists in itself (ontologically) but is unknown to the observer.
It only becomes known during the measurement.

The goal of the present paper is to present the Broglie-Bohm pilot-wave through
the study of the three most important experiments of quantum measurement:
the double-slit experiment which is the crucial experiment of the
wave-particle duality, the Stern and Gerlach experiment with the
measurement of the spin, and the EPR-B experiment with the problem
of non-locality.

The paper is organized as follows. In section II, we demonstrate the conditions in which the de Broglie-Bohm interpretation
can be assumed to be valid through continuity with classical mechanics. This involves the de Broglie-Bohm interpretation
for a set of particles prepared in the same way. In section~\ref{sect:DoubleSlit}, 
we present a numerical simulation of the double-slit experiment 
performed by Jönsson in 1961 with electrons~\cite{Jonsson}. 
The method of Feynman path integrals allows to calculate the time-dependent wave function. The evolution of the probability density 
just outside the slits leads one to consider the dualism of the 
wave-particle interpretation. And the de Broglie-Bohm trajectories 
provide an explanation for the impact positions of the particles. Finally, we show the continuity between 
classical and quantum trajectories with the convergence of these trajectories to classical 
trajectories when $h$ tends to $0$. In section~\ref{sect:SternGerlach}, we present an analytic expression of 
the wave function in the Stern-Gerlach experiment. This explicit solution 
requires the calculation of a Pauli spinor with a spatial extension.  
This solution enables to demonstrate the decoherence of the wave function and the three postulates of quantum measurement: 
quantization, Born interpretation and  wave function reduction. 
The spinor spatial extension also enables the introduction of the de
Broglie-Bohm trajectories which gives a very simple explanation of
the particles' impact and of the measurement process. In section~\ref{sect:EPR-B}, we study the EPR-B experiment, the Bohm version of the
Einstein-Podolsky-Rosen experiment. Its theoretical resolution in
space and time shows that a causal interpretation exists where each
atom has a position and a spin. Finally, we recall that a physical explanation of
non-local influences is possible.

\section{The de Broglie-Bohm interpretation}
\label{sect:deBroglieInterpretation}

The de Broglie-Bohm interpretation is based on the following demonstration.
Let us consider a wave function $\Psi(\textbf{x},t)$ solution to the Schr\"odinger
equation:
\begin{eqnarray}\label{eq:schrodinger1}
i\hbar \frac{\partial \Psi(\mathbf{x},t) }{\partial t}=\mathcal{-}\frac{\hbar ^{2}}{2m}%
\triangle \Psi(\mathbf{x},t) +V(\mathbf{x})\Psi(\mathbf{x},t)\\
\label{eq:schrodinger2}
\Psi (\mathbf{x},0)=\Psi_{0}(\mathbf{x}).
\end{eqnarray}
With the variable change $\Psi(\mathbf{x},t)=\sqrt{\rho^{\hbar}(\mathbf{x},t)} \exp(i\frac{S^{\hbar}(\textbf{x},t)}{\hbar})$, the Schr\"odinger
equation can be decomposed into Madelung
equations~\cite{Madelung_1926}(1926):
\begin{equation}\label{eq:Madelung1}
\frac{\partial S^{\hbar}(\mathbf{x},t)}{\partial t}+\frac{1}{2m}
(\nabla S^{\hbar}(\mathbf{x},t))^2 +
V(\mathbf{x})-\frac{\hbar^2}{2m}\frac{\triangle
\sqrt{\rho^{\hbar}(\mathbf{x},t)}}{\sqrt{\rho^{\hbar}(\mathbf{x},t)}}=0
\end{equation}
\begin{equation}\label{eq:Madelung2}
\frac{\partial \rho^{\hbar}(\mathbf{x},t)}{\partial t}+ div
\left(\rho^{\hbar}(\mathbf{x},t) \frac{\nabla
S^{\hbar}(\mathbf{x},t)}{m}\right)=0
\end{equation}
with initial conditions:
\begin{equation}\label{eq:Madelung3}
\rho^{\hbar}(\mathbf{x},0)=\rho^{\hbar}_{0}(\mathbf{x}) \qquad \text{and}
\qquad S^{\hbar}(\mathbf{x},0)=S^{\hbar}_{0}(\mathbf{x}) .
\end{equation}

Madelung equations correspond to a set of non-interacting quantum particles all prepared
in the same way (same  $\rho^{\hbar}_{0}(\mathbf{x})$
and $S^{\hbar}_{0}(\mathbf{x})$). 

A quantum particle is said to be \textit{statistically
prepared} if its initial probability density $\rho^{\hbar}_{0}(\mathbf{x})$
and its initial action $S^{\hbar}_{0}(\mathbf{x})$ converge, when $\hbar\to 0$, to non-singular functions $\rho_{0}(\mathbf{x})$ and $S_{0}(\mathbf{x})$. It is the case of an
electronic or $C_{60}$ beam in the double slit experiment or
an atomic beam in the Stern and Gerlach experiment. We will seen that it is also the case of a beam of entrangled particles in the EPR-B experiment.
 Then, we have the following theorem:~\cite{Gondran2011,Gondran2012a}

\textit{For statistically
prepared quantum particles,
the probability density} $\rho^{\hbar}(\textbf{x},t)$ \textit{and the
action} $S^{\hbar}(\textbf{x},t)$, \textit{solutions to the Madelung
equations
(\ref{eq:Madelung1})(\ref{eq:Madelung2})(\ref{eq:Madelung3}),
converge, when} $\hbar\to 0$,\textit{ to the classical density
$\rho(\textbf{x},t)$ and the classical action} $S(\textbf{x},t)$,
\textit{solutions to the statistical Hamilton-Jacobi equations:}
\begin{eqnarray}\label{eq:statHJ1b}
\frac{\partial S\left(\textbf{x},t\right) }{\partial
t}+\frac{1}{2m}(\nabla S(\textbf{x},t) )^{2}+V(\textbf{x},t)=0\\
\label{eq:statHJ2b}
S(\textbf{x},0)=S_{0}(\textbf{x})\\
\label{eq:statHJ3b}
\frac{\partial \mathcal{\rho }\left(\textbf{x},t\right) }{\partial
t}+ div \left( \rho \left( \textbf{x},t\right) \frac{\nabla
S\left( \textbf{x},t\right) }{m}\right) =0\\
\label{eq:statHJ4b}
\rho(\mathbf{x},0)=\rho_{0}(\mathbf{x}).
\end{eqnarray}

We give some indications on the demonstration of this theorem when the
wave function $\Psi(\textbf{x},t)$ is written as a
function of the initial wave function $\Psi_{0}(\textbf{x})$ by
the Feynman paths integral \cite{Feynman}:
\begin{equation}\label{eq:interFeynman}
\Psi(\textbf{x},t)= \int F(t,\hbar)
\exp\left(\frac{i}{\hbar}S_{cl}(\textbf{x},t;\textbf{x}_{0}\right)
\Psi_{0}(\textbf{x}_{0})d\textbf{x}_0
\end{equation}
where $F(t,\hbar)$ is an independent function of $\textbf{x}$ and
of $\textbf{x}_{0}$. 
For a statistically prepared quantum particle, the wave function is written
$ \Psi(\textbf{x},t)= F(t,\hbar)\int\sqrt{\rho^{\hbar}_0(\mathbf{x}_0)}
\exp(\frac{i}{\hbar}( S^{\hbar}_0(\textbf{x}_0)+
S_{cl}(\textbf{x},t;\textbf{x}_{0})) d\textbf{x}_0$. The theorem
of the stationary phase shows that, if $\hbar$ tends towards 0, we
have $ \Psi(\textbf{x},t)\sim
\exp(\frac{i}{\hbar}min_{\textbf{x}_0}( S_0(\textbf{x}_0)+
S_{cl}(\textbf{x},t;\textbf{x}_{0}))$, that is to say that the
quantum action $S^{h}(\textbf{x},t)$ converges to the function
\begin{equation}\label{eq:solHJminplus}
S(\textbf{x},t)=min_{\textbf{x}_0}( S_0(\textbf{x}_0)+
S_{cl}(\textbf{x},t;\textbf{x}_{0}))
\end{equation}
which is the solution to the Hamilton-Jacobi equation
(\ref{eq:statHJ1b}) with the initial condition (\ref{eq:statHJ2b}).
Moreover, as the quantum density $\rho^{h}(\textbf{x},t)$
satisfies the continuity equation (\ref{eq:Madelung2}), we deduce,
since $S^{h}(\textbf{x},t)$ tends towards $S(\textbf{x},t)$, that
$\rho^{h}(x,t)$ converges to the classical density
$\rho(\textbf{x},t)$, which satisfies the continuity equation
(\ref{eq:statHJ3b}). We obtain both announced convergences.

These statistical Hamilton-Jacobi equations (\ref{eq:statHJ1b},\ref{eq:statHJ2b},\ref{eq:statHJ3b},\ref{eq:statHJ4b}) correspond to a set of classical particles prepared in the same way (same $\rho_{0}(\mathbf{x})$ and $S_{0}(\mathbf{x})$). These classical particles are trajectories obtained in Eulerian representation with the velocity field $ \mathbf{v}\left( \mathbf{x,}t\right) =\frac{\mathbf{\nabla }S\left( \mathbf{%
x,}t\right) }{m} $, but the density and the action are not sufficient to describe it completely. To know its position at time $t$, it is necessary to know its initial position. Because  the Madelung equations converge to 
the statistical Hamilton-Jacobi equations, it is logical to do the same in quantum mechanics. We conclude that a \textit{ statistically prepared quantum particle} is not completely described by
its wave function. It is necessary to add this
initial position and an equation to define the evolution of this
position in the time. It is the de Brogglie-Bohm interpretation where the position is called the "hidden variable".

The two first postulates of quantum mechanics, describing the
quantum state and its evolution,~\cite{CT_1977} must be completed in this heterodox interpretation. 
At initial time t=0, the state of the particle is
given by the initial wave function $ \Psi_{0}(\textbf{x})$ (a wave
packet) and its initial position $\textbf{X}(0)$; it is the new first postulate.
The new second postulate gives the evolution on the wave function and on the position. 
For a single spin-less particle in a potential
$V(\textbf{x})$, the evolution of the wave function is given by the usual
Schrödinger equation (\ref{eq:schrodinger1},\ref{eq:schrodinger2})
and the evolution of the particle position is given by
\begin{equation}\label{eq:champvitesse}
\frac{d
\textbf{X}(t)}{dt}=\frac{\textbf{J}^{h}(\textbf{x},t)}{\rho^{h}(\textbf{x},t)}|_{\textbf{x}=\textbf{X}(t)}=\frac{\nabla
S^{h}(\textbf{x},t)}{m}|_{\textbf{x}=\textbf{X}(t)}
\end{equation}
where
\begin{equation}\label{eq:courant}
\textbf{J}^{h}(\textbf{x},t)=\frac{\hbar}{2 m i}
 ( \Psi^*(\textbf{x},t)\nabla\Psi(\textbf{x},t)-\Psi(\textbf{x},t)\nabla\Psi^*(\textbf{x},t))%=\rho(\textbf{x},t)\textbf{v}(\textbf{x},t)
\end{equation}
is the usual quantum current.

In the case of a particle with spin, as in the Stern and Gerlach
experiment, the Schrödinger equation must be replaced by the
Pauli or Dirac equations.

The third quantum mechanics postulate which describes the measurement 
operator (the observable) can be conserved. But the three postulates of 
measurement are not necessary: the postulate of quantization, the Born 
postulate of probabilistic interpretation of the wave function and the 
postulate of the reduction of the wave function. We see that
these postulates of measurement can be explained on each
example as we will shown in the following.

We replace these three postulates by a single one, the "quantum equilibrium 
hypothesis",~\cite{Durr_1992,Sanz,Norsen} that describes the interaction 
between the initial wave function $\Psi_0(\textbf{x})$ and the initial particle position
$\textbf{X}(0)$: 
For a set of identically prepared particles having $t=0$ 
wave function $\Psi_0(\textbf{x})$, it is assumed that
the initial particle positions $\textbf{X}(0)$ are distributed
according to:
\begin{equation}\label{eq:quantumequi}
P[\textbf{X}(0)=\textbf{x}]\equiv
P(\textbf{x},0)=|\Psi_0(\textbf{x})|^2 =\rho^{h}_0(\textbf{x}).
\end{equation}
It is the Born rule at the initial time.

Then, the probability distribution ($P(\textbf{x},t)\equiv
P[\textbf{X}(t)=\textbf{x}]$) for a set of particles moving
with the velocity field $\textbf{v}^{h}(\textbf{x},t)=\frac{\nabla
S^{h}(\textbf{x},t)}{m}$ satisfies the property of the "equivariance" of the
$|\Psi(\textbf{x},t)|^2$ probability distribution:~\cite{Durr_1992}
\begin{equation}\label{eq:quantumequit}
P[\textbf{X}(t)=\textbf{x}]\equiv
P(\textbf{x},t)= |\Psi(\textbf{x},t)|^2 =\rho^{h}(\textbf{x},t).
\end{equation}
It is the Born rule at time t.

Then, the de Broglie-Bohm interpretation is based on a continuity between classical and quantum mechanics where the quantum particles are statistical prepared with an initial probability densitiy satisfies the "quantum equilibrium 
hypothesis" (\ref{eq:quantumequi}). It is the case of the three studied experiments.

We will revisit these three measurement experiments through
mathematical calculations and numerical simulations. For each one,
we present the statistical interpretation that is common to the
Copenhagen interpretation and the de Broglie-Bohm pilot wave, then
the trajectories specific to the de Broglie-Bohm interpretation.
We show that the precise definition of the initial conditions,
i.e. the preparation of the particles, plays a fundamental
methodological role.

\section{Double-slit experiment with electrons}
\label{sect:DoubleSlit}

Young's double-slit experiment\cite{Young_1802} has long been the crucial experiment 
for the interpretation of the wave-particle duality. There have been realized with massive objects, such as 
electrons~\cite{Davisson,Jonsson}, 
neutrons~\cite{Halbon}, cold neutrons~\cite{Zeilinger_1988},
atoms~\cite{Estermann}, and more recently, with coherent ensembles
of ultra-cold atoms~\cite{Shimizu}, and even with
mesoscopic single quantum objects such as C$_{60}$ and
C$_{70}$~\cite{Arndt}. For Feynman, this experiment addresses 
"\emph{the basic element of the mysterious behavior [of electrons] in its most strange form. 
[It is] a phenomenon which is impossible, absolutely impossible to explain in any classical 
way and which has in it the heart of quantum mechanics. In reality, it contains the only 
mystery.}"~\cite{Feynman_1965}
The de Broglie-Bohm interpretation and the numerical simulation help us here to revisit the 
double-slit experiment with electrons performed by Jönsson in 1961 and to provide an answer to 
Feynman'mystery. These simulations~\cite{Gondran_2005a} follow those conducted in 1979 by Philippidis, Dewdney 
and Hiley~\cite{Philippidis_1979} which are today a classics. However, these simulations~\cite{Philippidis_1979} have some limitations because 
they did not consider realistic slits. The slits, which can be clearly represented by a function 
$G(y)$ with $G(y)=1$ for $-\beta\leq y \leq \beta$ and $G(y)=0$ for $|y|>\beta$, 
if they are $2\beta$ in width, were modeled by a Gaussian function $G(y)=e^{-y^2/2 \beta^2}$.
Interference was found, but the calculation could not account for
diffraction at the edge of the slits. Consequently, these simulations could not be used to defend the de Broglie-Bohm interpretation.

\begin{figure}[h]
\includegraphics[width=0.9\linewidth]{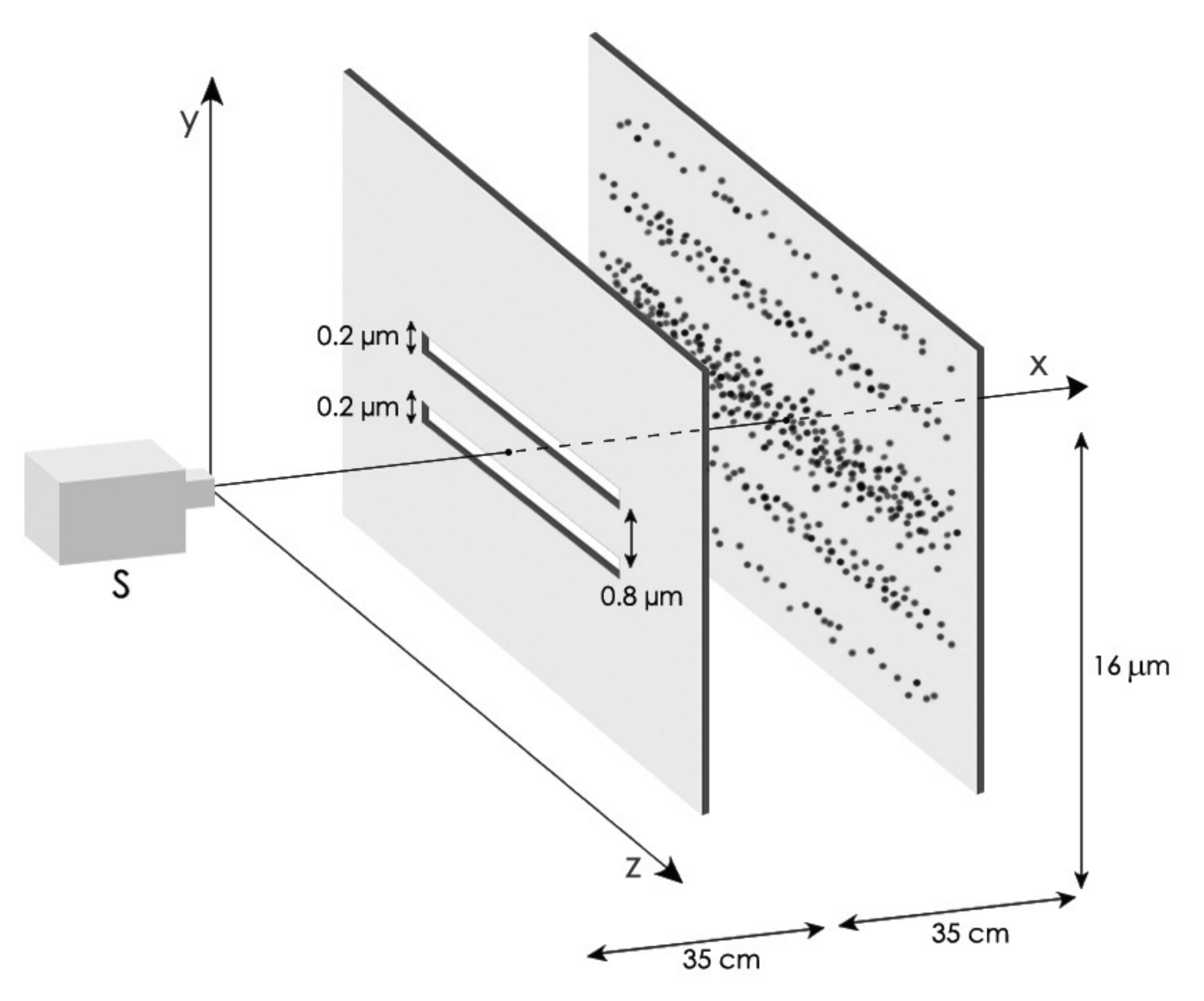}
\caption{\label{fig:schema-Young}Diagram of the Jönnson's double slit experiment performed with electrons.}
\end{figure}

Figure~\ref{fig:schema-Young} shows a diagram of the double slit experiment by Jönsson. 
An electron gun emits electrons one by one in the horizontal plane, through a hole 
of a few micrometers, at a velocity $v= 1.8\times10^{8} m/s$ along the horizontal $x$-axis. 
After traveling for $d_1= 35 cm$, they encounter a plate pierced with two 
horizontal slits A and B, each $0.2\mu$m wide and spaced $1\mu$m from each other. 
A screen located at $d_2=35cm$ after the slits collects these electrons. 
The impact of each electron appears on the screen as the experiment unfolds. 
After thousands of impacts, we find that the distribution of electrons on the 
screen shows interference fringes.

The slits are very long along the $z$-axis, so there is no effect of diffraction 
along this axis. In the simulation, we therefore only consider the wave function 
along the $y$-axis; the variable $x$ will be treated classically with $x=vt$. 
Electrons emerging from an electron gun are represented by the same initial 
wave function $\Psi_0(y)$.  

\subsection{Probability density}

\begin{figure}[h]
\includegraphics[width=0.8\linewidth]{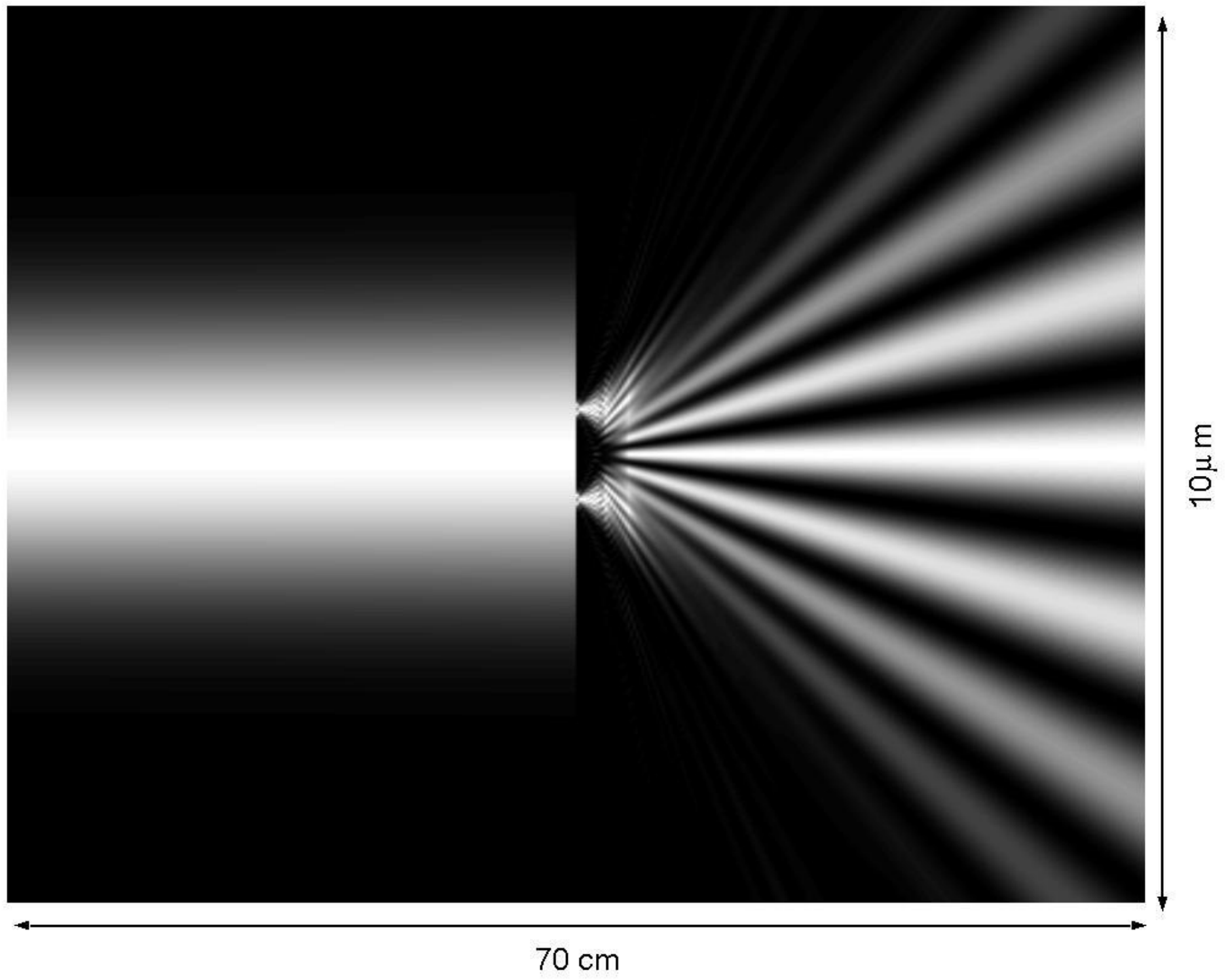}
\caption{\label{fig:schema-Young2}
General view of the evolution of the probability density from the source to the screen in the Jönsson experiment.
A lighter shade means that the density is higher i.e. the probability of presence is high.}
\end{figure}

Figure~\ref{fig:schema-Young2} gives a general view of the evolution of the probability 
density from the source to the detection screen (a lighter shade means that the density 
is higher i.e. the probability of presence is high). The calculations were made using 
the method of Feynman path integrals~\cite{Gondran_2005a}. The wave function after the slits ($t_1=d_1/v\simeq2. 10^{-11}s < t <t_1+d_2/v \simeq 4. 10^{-11}s$) is deduced 
from the values of the wave function at slits A and B: $\Psi(y,t)= \Psi_A(y,t)+ \Psi_B(y,t)$ 
with $\Psi_A(y,t)=\int_A K(y,t,y_a, t_1) \Psi(y_a, t_1) dy_a$, $\Psi_B(y,t)=\int_B 
K(y,t,y_b, t_1) \Psi(y_b, t_1) dy_b$ and $K(y,t,y_\alpha, t_1)= 
(m/2i\hbar (t-t_1))^{1/2} e^{im(y-y_\alpha)^2/2\hbar(t-t_1)}$.

\begin{figure}[h]
\includegraphics[width=0.8\linewidth]{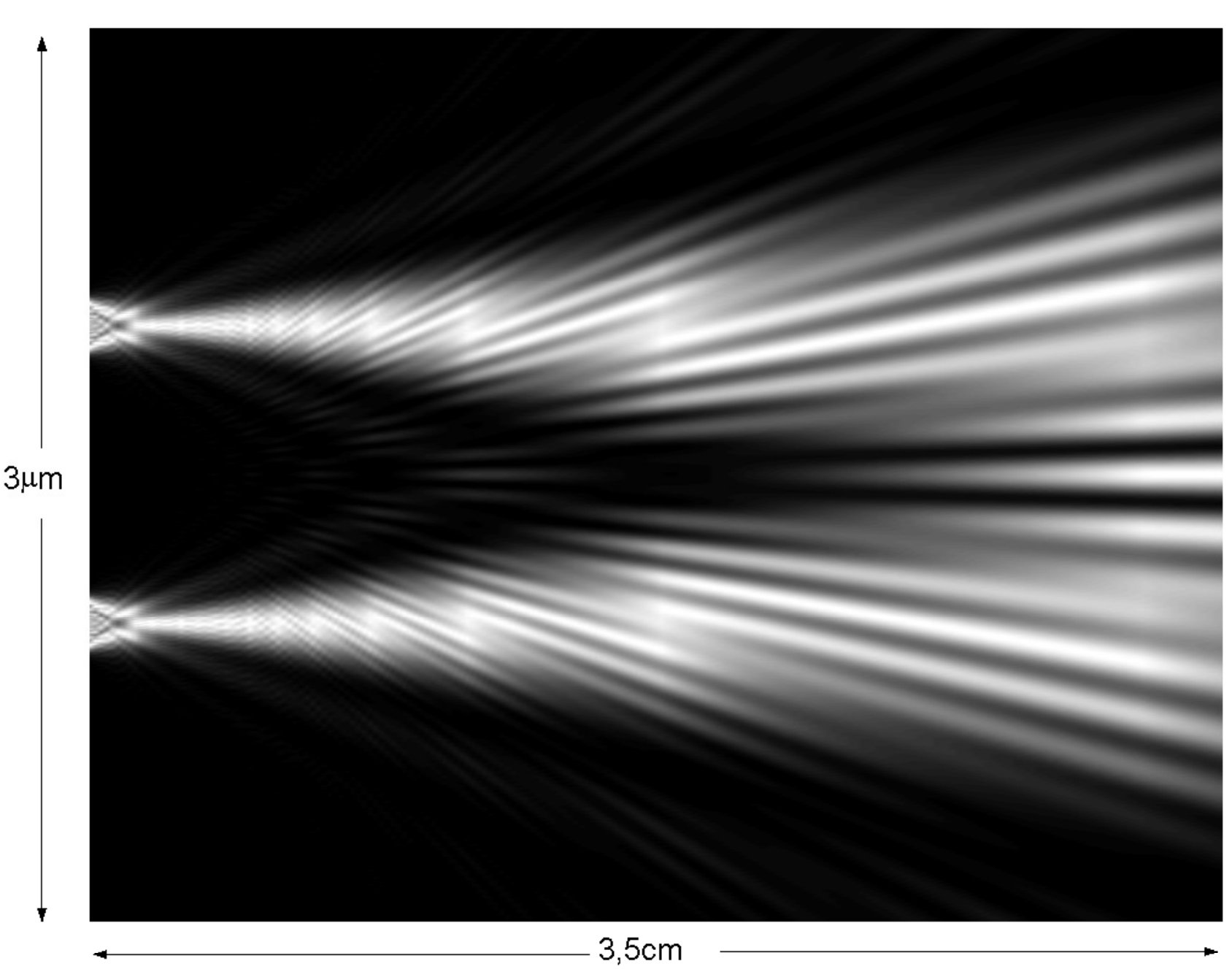}
\caption{\label{fig:schema-Young3}Close-up of the evolution of the probability density in the first $3cm$ after the slits in the Jönsson experiment.}
\end{figure}

Figure~\ref{fig:schema-Young3} shows a close-up of the evolution of the probability density 
just after the slits. We note that interference will only occur a few centimeters after the slits.
 Thus, if the detection screen is $1 cm$ from the slits, there is no interference and one can 
determine by which slit each electron has passed. In this experiment, the measurement is performed by the detection 
screen, which only reveals the existence or absence of the fringes.

\begin{figure}[h!]
\includegraphics[width=0.9\linewidth]{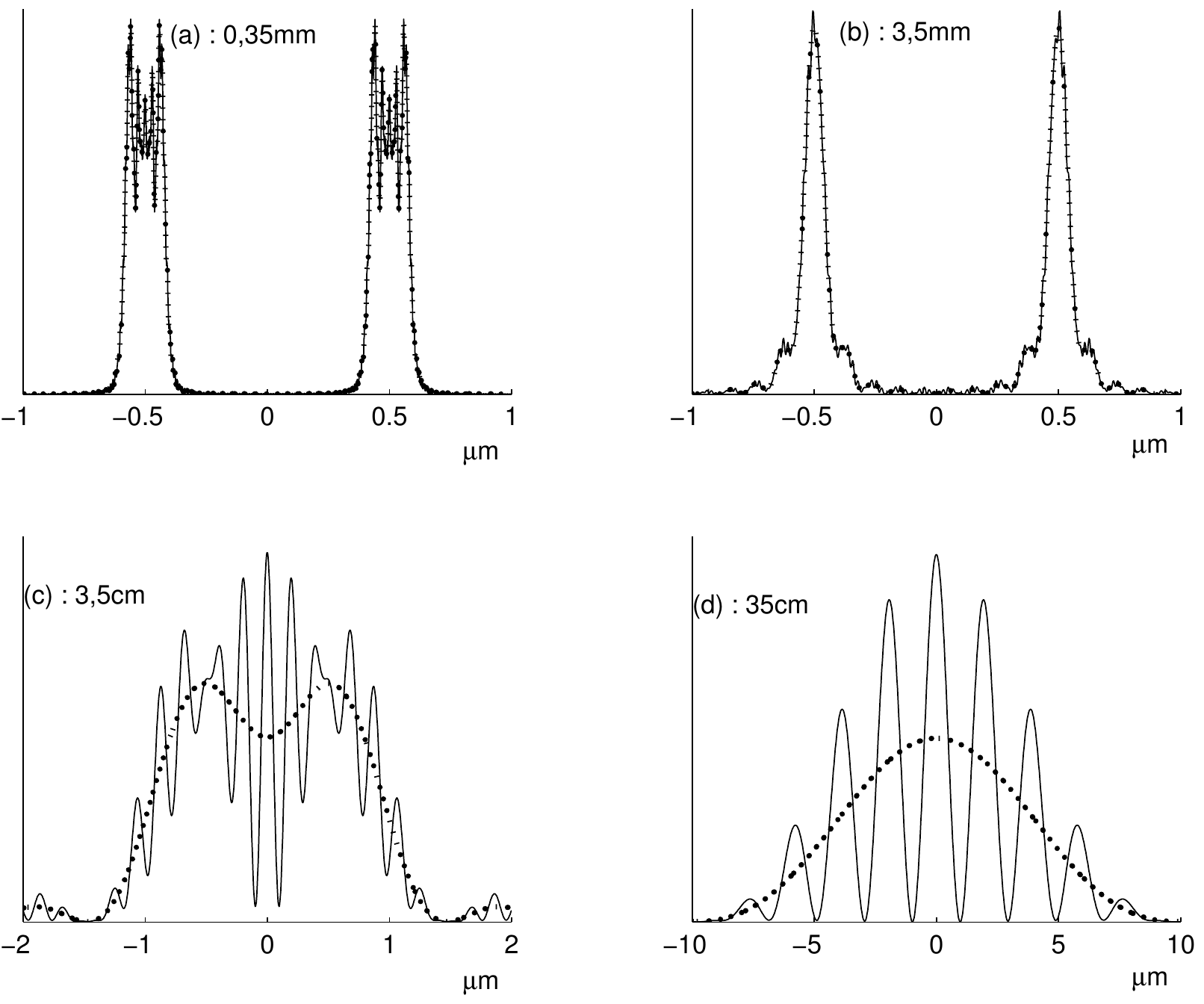}
\caption{\label{fig:schema-Young5}Comparison of the probability density $\vert \Psi_A +\Psi_B \vert^2$ (full line) 
and $\vert \Psi_A \vert^2+\vert\Psi_B \vert^2$ (dotted line) at various distances after the slits: 
(a) $0.35 mm$, (b): $3.5 mm$, (c): $3.5 cm$ and (d): $35 cm$.}
\end{figure}

The calculation method enables us to compare the evolution of the cross-section of the probability 
density at various distances after the slits ($0.35 mm$, $3.5 mm$, $3.5 cm$ and $35 cm$) where the 
two slits A and B are open simultaneously (interference: $ \vert \Psi_A +\Psi_B \vert^2$) with the 
evolution of the sum of the probability densities where the slits A and B are open independently 
(the sum of two diffractions: $ \vert \Psi_A \vert^2+\vert\Psi_B \vert^2$). Figure~\ref{fig:schema-Young5} 
shows that the difference between these two phenomena appears only a few centimeters after the slits.

\subsection{ Impacts on screen and de Broglie-Bohm trajectories}

The interference fringes are observed after a certain period of time when the impacts of the electrons 
on the detection screen become sufficiently numerous. Classical quantum theory only explains the impact 
of individual particles statistically.

However, in the de Broglie-Bohm interpretation: a particle has an initial position and follows a path whose 
velocity at each instant is given by equation (\ref{eq:champvitesse}). On the basis of this assumption we conduct 
a simulation experiment by drawing random initial positions of the electrons in the initial wave packet 
("quantum equilibrium hypothesis").

\begin{figure}[h]
\includegraphics[width=0.9\linewidth]{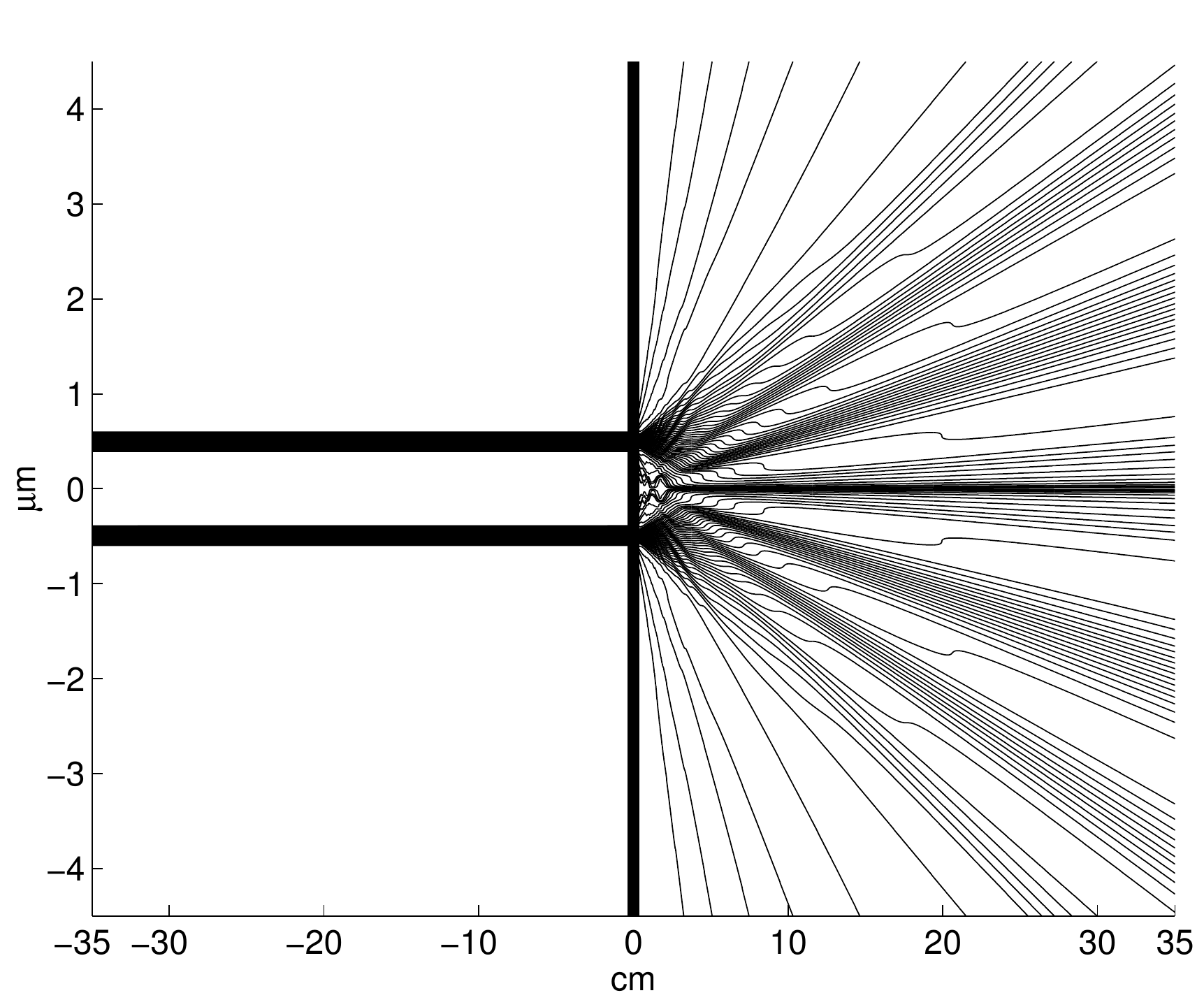}
\caption{\label{fig:traj-Young} 100 electron trajectories for the Jönsson experiment.}
\end{figure}

\begin{figure}[h]
\includegraphics[width=0.9\linewidth]{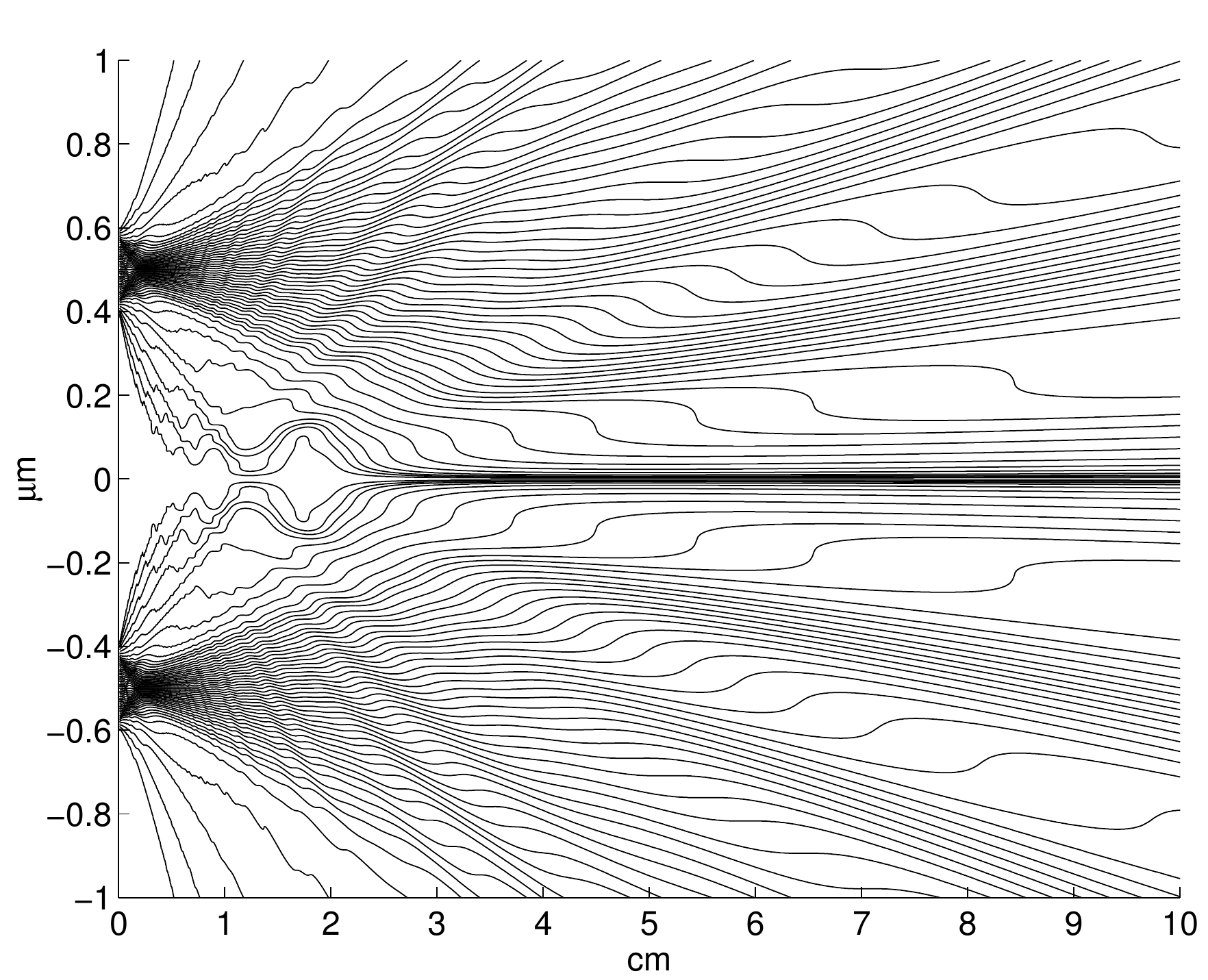}
\caption{\label{fig:zoom-traj-Young} Close-up on the 100 trajectories of the electrons just after the slits.}
\end{figure}

Figure~\ref{fig:traj-Young} shows, after its initial starting position, 100 possible quantum 
trajectories of an electron passing through one of the two slits: We have not represented the paths 
of the electron when it is stopped by the first screen. Figure~\ref{fig:zoom-traj-Young} shows 
a close-up of these trajectories just after they leave their slits.

The different trajectories explain both the impact of electrons on the detection screen and the interference fringes.
This is the simplest and most natural interpretation to explain the impact positions: "The position of an impact is 
simply the position of the particle at the time of impact." This was the view defended by Einstein at the Solvay 
Congress of 1927. The position is the only measured variable of the experiment.

In the de Broglie-Bohm interpretation, the impacts on the screen are the real positions of the electron as in classical mechanics and the three postulates of the measurement of quantum mechanics can be trivialy explained: the position is an eigenvalue of the position operator because the position variable is identical to its operator (X$\Psi $ = x$\Psi $), the Born postulate is satisfied with the "equivariance" property, and the reduction of the wave packet is not necessary 
to explain the impacts.

\begin{figure}[h]
\includegraphics[width=0.9\linewidth]{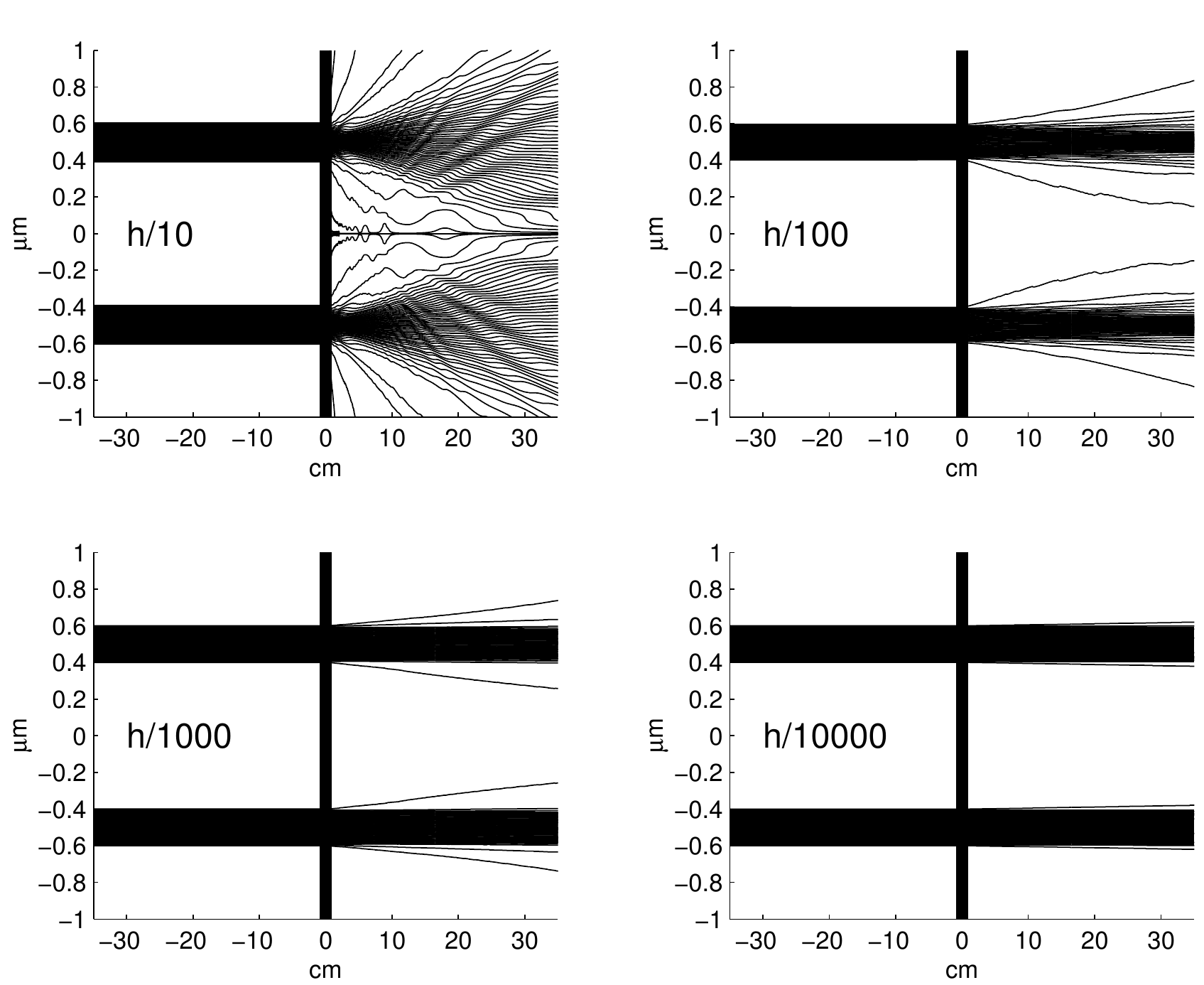}
\caption{\label{fig:traj-Young-converg}Convergence of 100 electron trajectories when h is divided by 10, 100, 1 000 and 10 000.}
\end{figure}

Through numerical simulations, we will demonstrate how, when the Planck constant $h$ tends to 0, 
the quantum trajectories converge to the classical trajectories. In reality a constant is not able to tend to 0 by definition. The convergence to classical trajectories 
is obtained if the term $ht/m\rightarrow0$; so $h\rightarrow0$ is equivalent to $m\rightarrow+\infty$ 
(i.e. the mass of the particle grows) or $t\rightarrow0$ (i.e. the distance slits-screem $d_2\rightarrow0$). 
Figure~\ref{fig:traj-Young-converg} shows the 100 trajectories that start at the same 100 initial 
points when Planck's constant is divided respectively by 10, 100, 1000 and 10000 (equivalent to multiplying the mass by 10, 100, 1000 and 10000). 
We obtain quantum trajectories converging to the classical trajectories, when $h$ tends to 0.

The study of the slits clearly shows that, in the de Broglie-Bohm interpretation, there is no physical separation between quantum mechanics and classical mechanics. All particles have quantum properties, but specifically quantum behavior only appears in certain experimental conditions: here when the ratio ht/m is sufficiently large. Interferences only appear gradually and the quantum particle behaves at any time as both a wave and a particle.

\section{The Stern-Gerlach experiment}
\label{sect:SternGerlach}

\begin{figure}[h]
\includegraphics[width=0.9\linewidth]{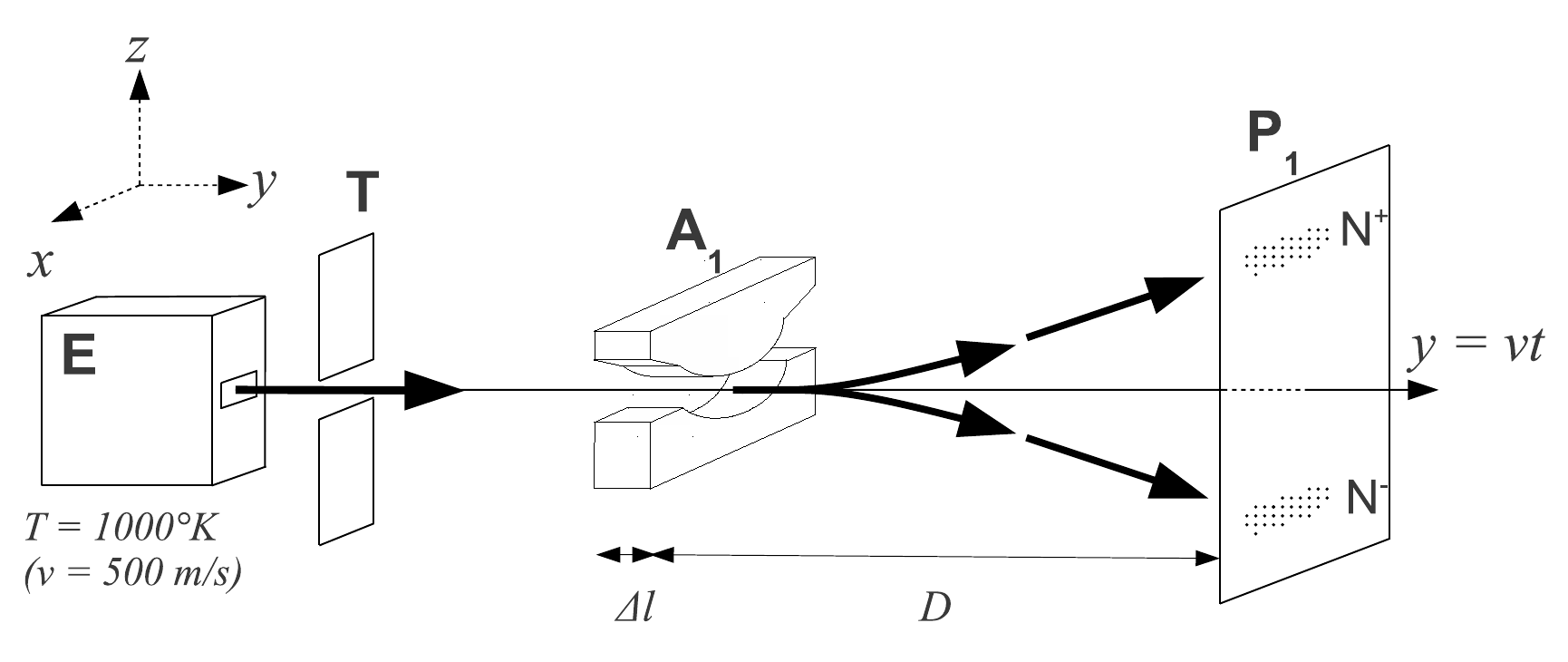}
\caption{\label{fig:schema-SetG}Schematic configuration of the
Stern-Gerlach experiment.}
\end{figure}

In 1922, by studying the deflection of a beam of silver atoms in a
strongly inhomogeneous magnetic field (cf.
Figure~\ref{fig:schema-SetG}) Otto Stern and Walter Gerlach
\cite{SternGerlach} obtained an experimental result that
contradicts the common sense prediction: the beam, instead of
expanding, splits into two separate beams giving two spots of
equal intensity $N^+$ and $N^-$ on a detector, at equal
distances from the axis of the original beam.

Historically, this is the experiment which helped establish
spin quantization. Theoretically, it is the seminal experiment
posing the problem of measurement in quantum mechanics. Today it
is the theory of decoherence with the diagonalization of the
density matrix that is put forward to explain the first part of
the measurement process \cite{Zeh}.
However, although these authors consider the Stern-Gerlach
experiment as fundamental, they do not propose a calculation of the spin decoherence time.

We present an analytical solution to this
decoherence time and the diagonalization of the density matrix.
This solution requires the calculation of the Pauli spinor with a
spatial extension as the equation:
\begin{equation}\label{eq:psi-0}
    \Psi^{0}(z) = (2\pi\sigma_{0}^{2})^{-\frac{1}{2}}
                      e^{-\frac{z^2}{4\sigma_0^2}}
                      \left( \begin{array}{c}\cos \frac{\theta_0}{2}e^{ - i\frac{\varphi_0}{2}}
                                   \\
                                  \sin\frac{\theta_0}{2}e^{i\frac{\varphi_0}{2}}
                  \end{array}
           \right).
\end{equation}
Quantum mechanics textbooks \cite{Feynman_1965, CT_1977, Sakurai,
LeBellac} do not take into account the spatial extension of the
spinor (\ref{eq:psi-0}) and simply use the simplified spinor
without spatial extension:
\begin{equation}\label{eq:psi-s}
    \Psi^{0} = \left( \begin{array}{c}\cos \frac{\theta_0}{2}e^{ - i\frac{\varphi_0}{2}}
                                   \\
                                  \sin\frac{\theta_0}{2}e^{i\frac{\varphi_0}{2}}
                  \end{array}
           \right).
\end{equation}
However, as we shall see, the different evolutions of the spatial
extension between the two spinor components will have a key role
in the explanation of the measurement process. This spatial
extension enables us, in following the precursory works of
Takabayasi~\cite{Takabayasi_1954}, Bohm~\cite{Bohm_1955,Bohm_1993},
Dewdney et al.~\cite{Dewdney_1986} and Holland~\cite{Holland_1993}, 
to revisit the Stern and Gerlach experiment, to explain the decoherence and to demonstrate
the three postulates of the measure: quantization, Born statistical 
interpretation and wave function reduction.

\begin{figure}[h]
\includegraphics[width=0.3\linewidth]{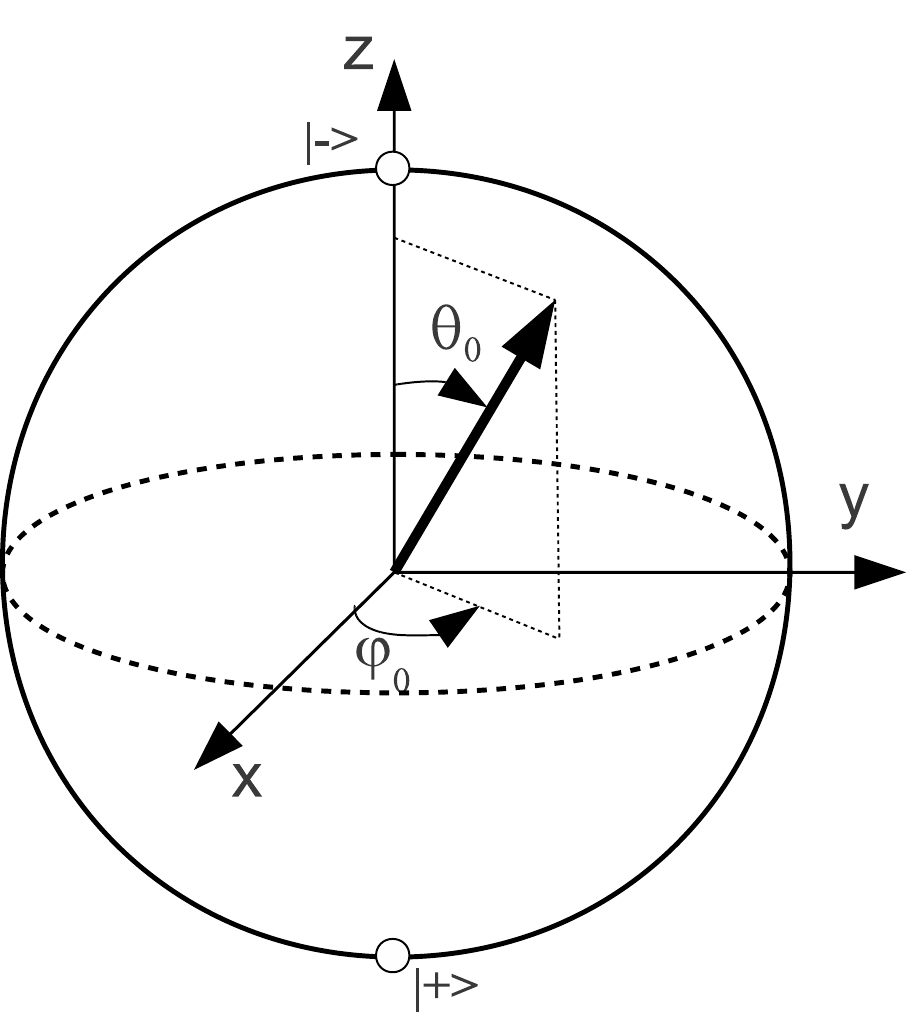}
\caption{\label{fig:spin}Orientation of the magnetic moment.
$\theta_0$ and $\varphi_{0}$ are the polar angles characterizing the spin 
vector in the de Broglie-Bohm interpretation.}
\end{figure}

Silver atoms contained in the oven E (Figure~\ref{fig:schema-SetG})
are heated to a high temperature and escape through a narrow
opening. A second aperture, T, selects those atoms whose velocity,
$\textbf{v}_0$, is parallel to the $y$-axis. The atomic beam crosses
the gap of the electromagnet $A_{1}$ before condensing on the
detector, $P_{1}$ . Before crossing the electromagnet, the
magnetic moment of each silver atom is oriented randomly
(isotropically). In the beam, we represent each atom by its wave
function; one can assume that at the entrance to the
electromagnet, $A_{1}$, and at the initial time $t=0$, each atom
can be approximatively described by a Gaussian spinor in $z$ given
by (\ref{eq:psi-0}) corresponding to a pure state. The variable $y$
will be treated classically with $y= vt$. $\sigma_0=10^{-4}m$
corresponds to the size of the slot T along the $z$-axis. The
approximation by a Gaussian initial spinor will allow explicit
calculations. Because the slot is much wider along the $x$-axis, the
variable $x$ will be also treated classically. 
To obtain an explicit solution of the Stern-Gerlach experiment, we take
the numerical values used in the CohenTannoudji 
textbook~\cite{CT_1977}. For the silver atom,
we have $m = 1.8\times 10^{-25} kg$, $v_0 = 500\ m/s$ 
(corresponding to the temperature of $T=1000°K$). 
In equation~(\ref{eq:psi-0}) and in figure~\ref{fig:spin}, 
$\theta_0$ and $\varphi_{0}$ are the polar angles characterizing the initial orientation of the
magnetic moment, $\theta_0$ corresponds to the angle with the
$z$-axis. The experiment is a statistical mixture of pure states
where the $\theta_0$ and the $\varphi_0$ are randomly chosen:
$\theta_0$ is drawn in a uniform way from $[0,\pi]$ and that
$\varphi_0$ is drawn in a uniform way from $[0,2\pi]$.

The evolution of the spinor $\Psi=\left( \begin{array}{c}\psi_{+}
                                 \\
                        \psi_{-}
                  \end{array}
           \right)$ in a magnetic field
$\textbf{B}$ is then given by the Pauli equation:
\begin{equation}\label{eq:Pauli}
    i\hbar \left( \begin{array}{c} \frac{\partial \psi _{+}}{\partial t}
                                   \\
                                   \frac{\partial \psi _{-}}{\partial t}
                  \end{array}
           \right)
    =-\frac{\hbar ^{2}}{2m} \Delta
                           \left( \begin{array}{c} \psi _{+}
                                                   \\
                                                   \psi _{-}
                                  \end{array}
                           \right)
     +\mu _{B}\textbf{B}\sigma \left( \begin{array}{c} \psi _{+}
                                                                  \\
                                                                  \psi _{-}
                                                 \end{array}
                                          \right)
\end{equation}
where $\mu_B=\frac{e\hbar}{2m_e}$ is the Bohr magneton and where
$\sigma=(\sigma_{x},\sigma_{y},\sigma_{z})$ corresponds to the
three Pauli matrices. The particle first enters an electromagnetic
field $\textbf{B}$ directed along the $z$-axis, $B_{x}=B'_0x$,
$B_{y}=0$, $B_{z}=B_{0} -B'_{0} z$, with $B_{0}=5$ Tesla,
$B'_{0}=\left| \frac{\partial B}{\partial z}\right| = 10^3~Tesla/m$ 
over a length $\Delta l=1~cm$. On exiting the magnetic
field, the particle is free until it reaches the detector $P_1$
placed at a $D=20~cm$ distance.

The particle stays within the magnetic field for a time $\Delta
t=\frac{\Delta l}{v}= 2\times 10^{-5} s$. During this time
$[0,\Delta t]$, the spinor is:~\cite{Platt_1992} (see Appendix A)
\begin{widetext}
\begin{equation}\label{eq:fonctiondanschampmagnétique}
\Psi (z,t)  \simeq \left(
\begin{array}{c}
                                \cos \frac{\theta_0}{2}
                 (2\pi\sigma_0^2)^{-\frac{1}{2}}
                 e^{-\frac{(z-\frac{\mu_{B} B'_{0}}{2 m}t^{2})^2 }
                 {4\sigma_0^2}} e^{i\frac{\mu_{B} B'_{0}t z -\frac{\mu^2_{B} B'^2_{0}}{6 m}t^3 +  \mu_B B_0 t + \frac{\hbar \varphi_0}{2}}{\hbar }}\\
                                i \sin \frac{\theta_0}{2}
                 (2\pi\sigma_0^2)^{-\frac{1}{2}}
                 e^{-\frac{(z+\frac{\mu_{B} B'_{0}}{2 m}t^{2})^2}
                 {4\sigma_0^2}} e^{i\frac{-\mu_B B'_{0}t z -\frac{\mu^2_{B} B'^2_{0}}{6
    m}t^3 -  \mu_B B_0 t -\frac{\hbar \varphi_0}{2}}{\hbar }}
                            \end{array}
                     \right).
\end{equation}
\end{widetext}

After the magnetic
field, at time $t+ \Delta t$ $(t \geq 0)$ in the free space, the
spinor becomes: \cite{Bohm_1993,Dewdney_1986,Holland_1993,Platt_1992,Gondran_2005b} (see Appendix A)
{\small
\begin{equation}\label{eq:fonctionapreschampmagnétique}
\Psi (z,t+\Delta t) \simeq\left(
\begin{array}{c}
                                \cos \frac{\theta_0}{2}
                 (2\pi\sigma_0^2)^{-\frac{1}{2}}
                 e^{-\frac{(z-z_{\Delta}- ut)^2 }
                 {4\sigma_0^2}} e^{i\frac{m u z + \hbar \varphi_+}{\hbar }} \\
                                \sin \frac{\theta_0}{2}
                (2\pi\sigma_0^2)^{-\frac{1}{2}}
                e^{-\frac{(z+z_{\Delta}+ ut)^2}
                {4\sigma_0^2}} e^{i\frac{-
    muz + \hbar \varphi_-}{\hbar }}
                            \end{array}
                     \right)
\end{equation}}
where
\begin{equation}\label{eq:zdeltavitesse}
    z_{\Delta}=\frac{\mu_B B'_{0}(\Delta
    t)^{2}}{2 m}=10^{-5}m,~~~~~~u =\frac{\mu_B B'_{0}(\Delta t)}{m}=1 m/s.
\end{equation}
Equation (\ref{eq:fonctionapreschampmagnétique}) takes into
account the spatial extension of the spinor and we note that the
two spinor components have very different $z$ values. All
interpretations are based on this equation.

\subsection{The decoherence time}

\begin{figure}[h]
\includegraphics[width=0.95\linewidth]{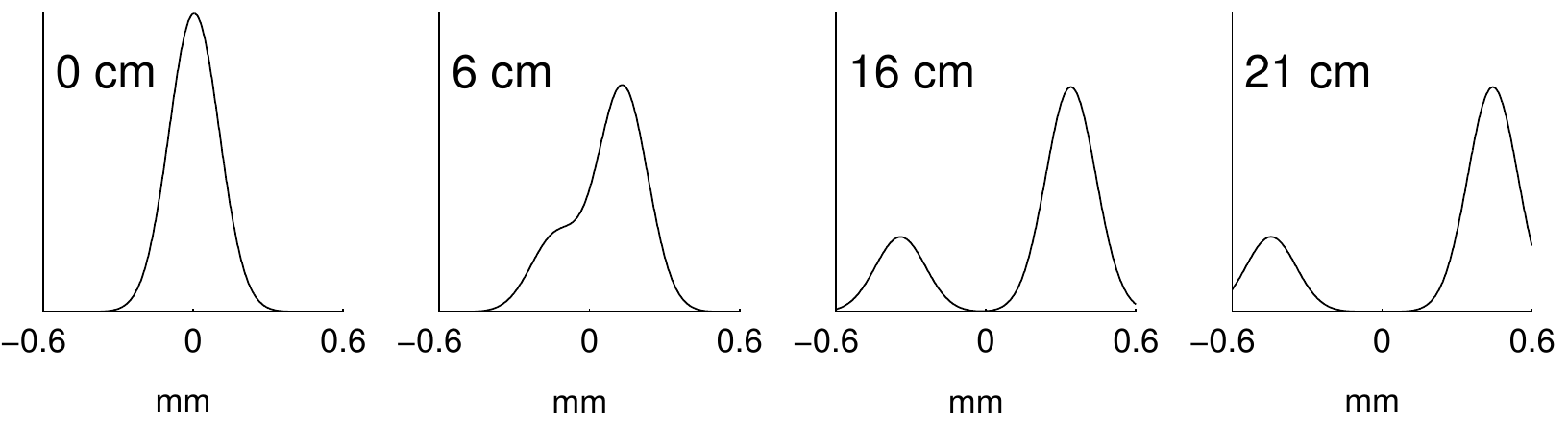}
\caption{\label{fig:ddp-SetG}Evolution of the probability density
of a pure state with $\theta_0= \pi/3$.}
\end{figure}

We deduce from (\ref{eq:fonctionapreschampmagnétique}) the
probability density of a pure state in the free space after the
electromagnet:
{\small
% \begin{eqnarray}\label{eq:densitéaprèschampmagnétiqueteta}
%     \rho_{\theta_0}(z,t+ \Delta t) &\simeq
%      (2\pi\sigma_0^2)^{-\frac{1}{2}}
%                   \left(\cos^{2} \frac{\theta_0}{2}  e^{-\frac{(z-z_{\Delta}- ut)^2}{2\sigma_0^2}}+
%                    \sin^{2} \frac{\theta_0}{2}
%                   e^{-\frac{(z+z_{\Delta}+ ut)^2}{2\sigma_0^2}}\right)
% \end{eqnarray}
\begin{eqnarray}
    \rho_{\theta_0}(z,t+ \Delta t) \simeq
     (2\pi\sigma_0^2)^{-\frac{1}{2}}
                  &&\left(\cos^{2} \frac{\theta_0}{2}  e^{-\frac{(z-z_{\Delta}- ut)^2}{2\sigma_0^2}}\right.\nonumber\\
                        &&\quad +\left.\sin^{2} \frac{\theta_0}{2}
                  e^{-\frac{(z+z_{\Delta}+ ut)^2}{2\sigma_0^2}}\right)\label{eq:densitéaprèschampmagnétiqueteta}
\end{eqnarray}}
Figure~\ref{fig:ddp-SetG} shows the probability density of a pure
state (with $\theta_0= \pi/3$) as a function of $z$ at several
values of $t$ (the plots are labeled $y = vt$). The beam separation
does not appear at the end of the magnetic field ($1~cm$), but $16~cm$
further along. It is the moment of the decoherence.
The decoherence time, where the two spots $N^{+}$ and $N^{-}$ are
separated, is then given by the equation:
\begin{equation}\label{eq:tempsdecoherence}
t_{D} \simeq \frac{3 \sigma_{0}-z_\Delta}{u}=3 \times
10^{-4}s.
\end{equation}

This decoherence time is usually the time required to diagonalize
the marginal density matrix of spin variables associated with a
pure state \cite{Roston}:
\begin{equation}\label{eq:matricedensite1}
\rho^{S}(t) =\left(
\begin{array}{ccccc}
  \int|\psi_+(z,t)|^2 dz &    \int\psi_+(z,t)\psi^*_-(z,t) dz\\
 \int\psi_-(z,t)\psi^*_+(z,t) dz  &   \int|\psi_-(z,t)|^2 dz\\
\end{array}
\right)
\end{equation}

For $t\geq t_D$, the product $\psi_+(z,t+ \Delta t)\psi_-(z,t+
\Delta t) $ is null and the density matrix is diagonal: the
probability density of the initial pure state
(\ref{eq:fonctionapreschampmagnétique}) is diagonal:
\begin{equation}\label{eq:matricedensite3}
\rho^{S}(t+ \Delta t) =  (2\pi\sigma_0^2)^{-1}\left(
\begin{array}{ccccc}
\cos^2\frac{\theta_0}{2}   &  0\\
0  & \sin^2\frac{\theta_0}{2}  \\
\end{array}
\right)
\end{equation}

\subsection{Proof of the postulates of quantum measurement}

\begin{figure}[h]
\includegraphics[width=0.8\linewidth]{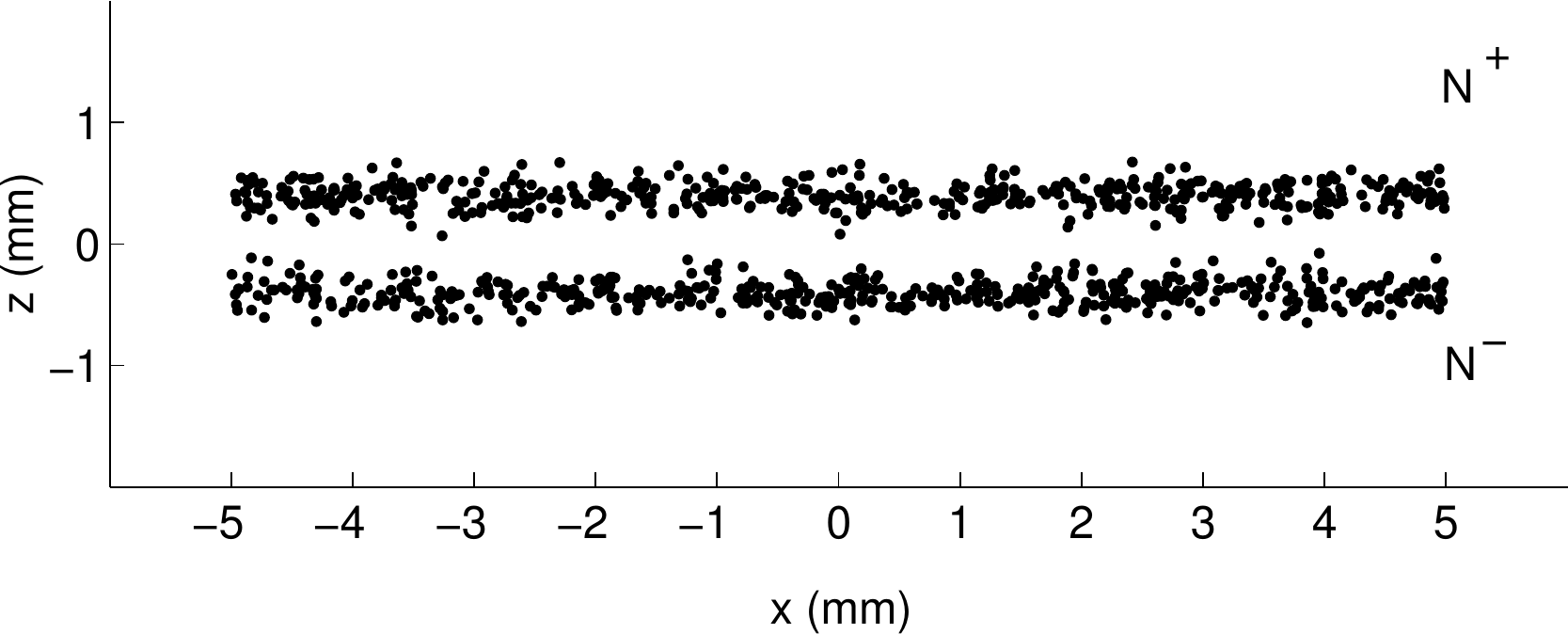}
\caption{\label{fig:impacts-SetG}1000 silver atom impacts on the detector $P_1$.}
\end{figure}

We then obtain atoms with a spin oriented only along the $z$-axis (positively
or negatively). Let us consider the spinor $\Psi(z,t+ \Delta t) $
given by equation~(\ref{eq:fonctionapreschampmagnétique}).
Experimentally, we do not measure the spin directly, but the
$\widetilde{z}$ position of the particle impact on $P_1$ (Figure~\ref{fig:impacts-SetG}).

If \ $\widetilde{z}\in N^+$, the term $\psi_-$ of
(\ref{eq:fonctionapreschampmagnétique}) is numerically equal to zero and the
spinor $\Psi$ is proportional to $\binom{1}
                            {0} $, one of the eigenvectors of the spin operator $S_z= \frac{\hbar}{2} \sigma_z $: $ \Psi (\tilde{z},t+\Delta t) \simeq
(2\pi\sigma_0^2)^{-\frac{1}{4}} \cos \frac{\theta_0}{2}
                 e^{-\frac{(\tilde{z}_1-z_{\Delta}- ut)^2 }
                 {4\sigma_0^2}} e^{i\frac{m u \tilde{z}_1 + \hbar \varphi_+}{\hbar }}\binom{1}
                            {0}$. Then, we have $S_z\Psi  =\frac{\hbar}{2} \sigma_z \Psi= +\frac{\hbar}{2} \Psi$.

If $\widetilde{z}\in N^-$, the term $\psi_+$ of
(\ref{eq:fonctionapreschampmagnétique}) is numerically equal to zero and the
spinor $\Psi $ is proportional to $\binom{0}
                            {1}$, the other eigenvector of the spin operator $S_z$: $ \Psi (\tilde{z},t+\Delta t) \simeq
(2\pi\sigma_0^2)^{-\frac{1}{4}}\sin \frac{\theta_0}{2}
                e^{-\frac{(\tilde{z}_2+z_{\Delta}+ ut)^2}
                {4\sigma_0^2}} e^{i\frac{-
    mu\tilde{z}_2 + \hbar \varphi_-}{\hbar }}\binom{0}
                            {1}$. Then, we have $S_z\Psi  =\frac{\hbar}{2} \sigma_z \Psi= -\frac{\hbar}{2} \Psi$.
Therefore, the measurement of the spin corresponds to an eigenvalue of the spin operator.                            
It is a proof of the postulate of quantization.

Equation (\ref{eq:matricedensite3}) gives the probability $
\cos^{2} \frac{\theta_0}{2} $ (resp. $\sin^{2} \frac{\theta_0}{2}
$) to measure the particle in the spin state $+ \frac{\hbar}{2}$
(resp. $-\frac{\hbar}{2}$); this proves the Born probabilistic postulate.

By drilling a hole in the detector $P_1$ to the location of the
spot $ N^+$ (figure~\ref{fig:schema-SetG}), we select all the atoms that are in the spin
state $ |+\rangle= \binom{1}
                            {0}$. The new spinor of these atoms is obtained by
making the component $\Psi_-$ of the spinor $\Psi$ identically
zero (and not only numerically equal to zero) at the time when the atom crosses the detector $P_1 $; at
this time the component $\Psi_-$ is indeed stopped by detector $
P_1 $. The future trajectory of the silver atom after crossing the detector $
P_1 $ will be guided by this new (normalized) spinor. The
wave function reduction is therefore not linked to the
electromagnet, but to the detector $P_1$ causing an irreversible
elimination of the spinor component $ \Psi_-$.

\subsection{Impacts and quantizations explained by de Broglie-Bohm trajectories}

Finally, it remains to provide an explanation of the individual
impacts of silver atoms. The spatial extension of the spinor
(\ref{eq:psi-0}) allows to take into account the particle's
initial position $z_0$ and to introduce the Broglie-Bohm
trajectories
\cite{Dewdney_1986,Holland_1993,Broglie_1927,Bohm_1952,Challinor}
which is the natural assumption to explain the individual
impacts.

\begin{figure}[H]
\begin{center}
\includegraphics[width=0.9\linewidth]{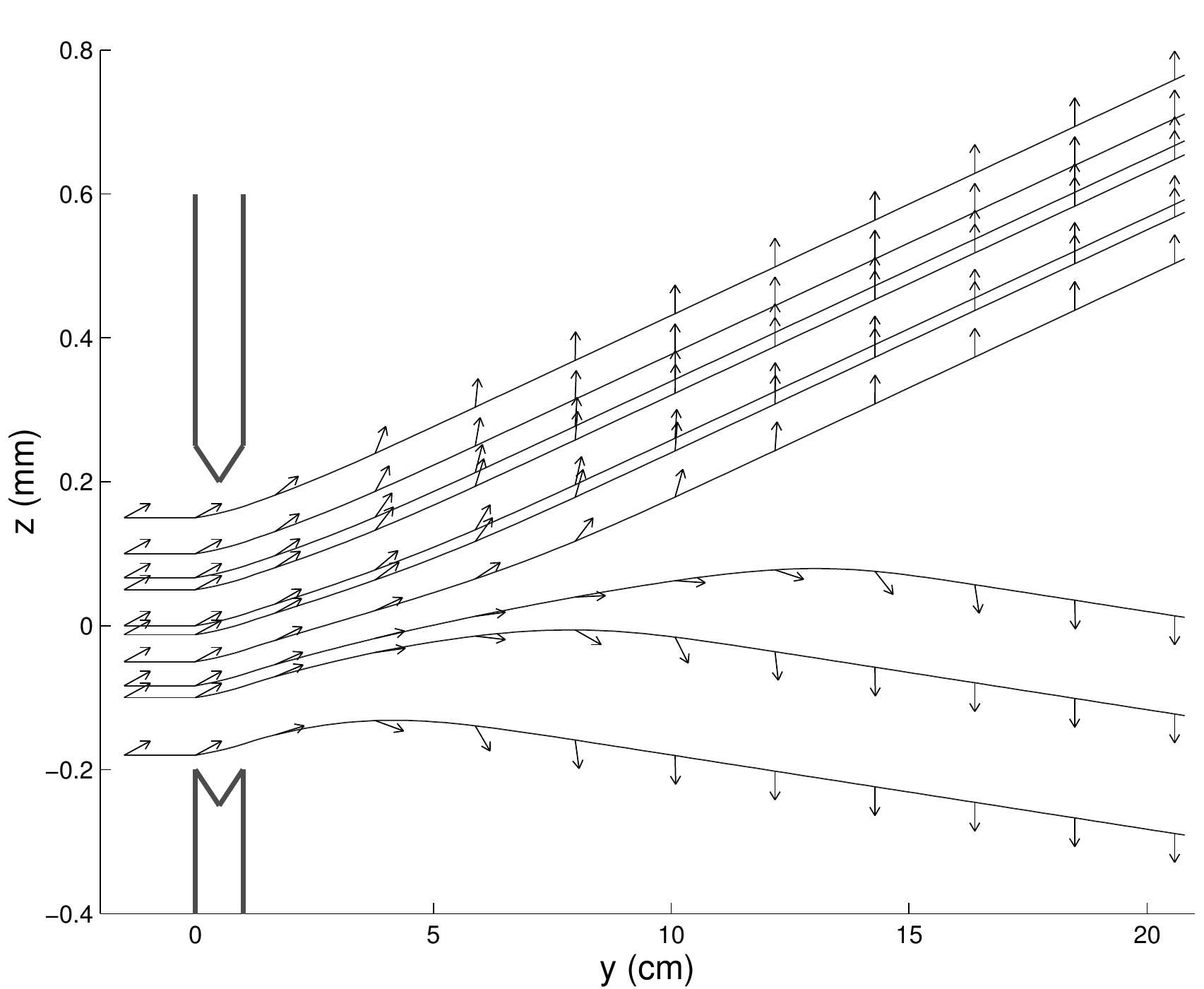}
\caption{\label{fig:SetG-10traj} Ten silver atom trajectories with
initial spin orientation $(\theta_0=\frac{\pi}{3})$ and initial
position $z_0$; arrows represent the spin orientation
$\theta(z,t)$ along the trajectories.}
\end{center}
\end{figure}

Figure~\ref{fig:SetG-10traj} presents, for a silver atom with the
initial spinor orientation $(\theta_0=\frac{\pi}{3},\varphi_0=0)$, a
plot in the $(Oyz)$ plane of a set of 10 trajectories whose initial
position $z_0$ has been randomly chosen from a Gaussian
distribution with standard deviation $\sigma_{0}$. The spin
orientations $\theta(z,t)$ are represented by arrows.

The final orientation, obtained after the decoherence time
$t_{D}$, depends on the initial particle position $z_{0}$ in the
spinor with a spatial extension and on the initial angle
$\theta_{0}$ of the spin with the $z$-axis. We obtain
$+\frac{\pi}{2}$ if $z_0
> z^{\theta_{0}}$ and $-\frac{\pi}{2}$ if $z_0
< z^{\theta_{0}}$ with
\begin{equation}\label{eq:seuilpolarization}
z^{\theta_{0}}=\sigma_0 F^{-1}\left(\sin^{2}\frac{\theta_{0}}{2}\right)
\end{equation}
where F is the repartition function of the normal centered-reduced
law. If we ignore the position of the atom in its wave function,
we lose the determinism given by equation
(\ref{eq:seuilpolarization}).

In the de Broglie-Bohm interpretation with a realistic interpretation of the spin, the "measured" value
is not independent of the context of the measure and is
contextual. It conforms to the Kochen and Specker
theorem:~\cite{Kochen 1967} Realism and non-contextuality are
inconsistent with certain quantum mechanics predictions. 

\begin{figure}[H]
\begin{center}
\includegraphics[width=0.9\linewidth]{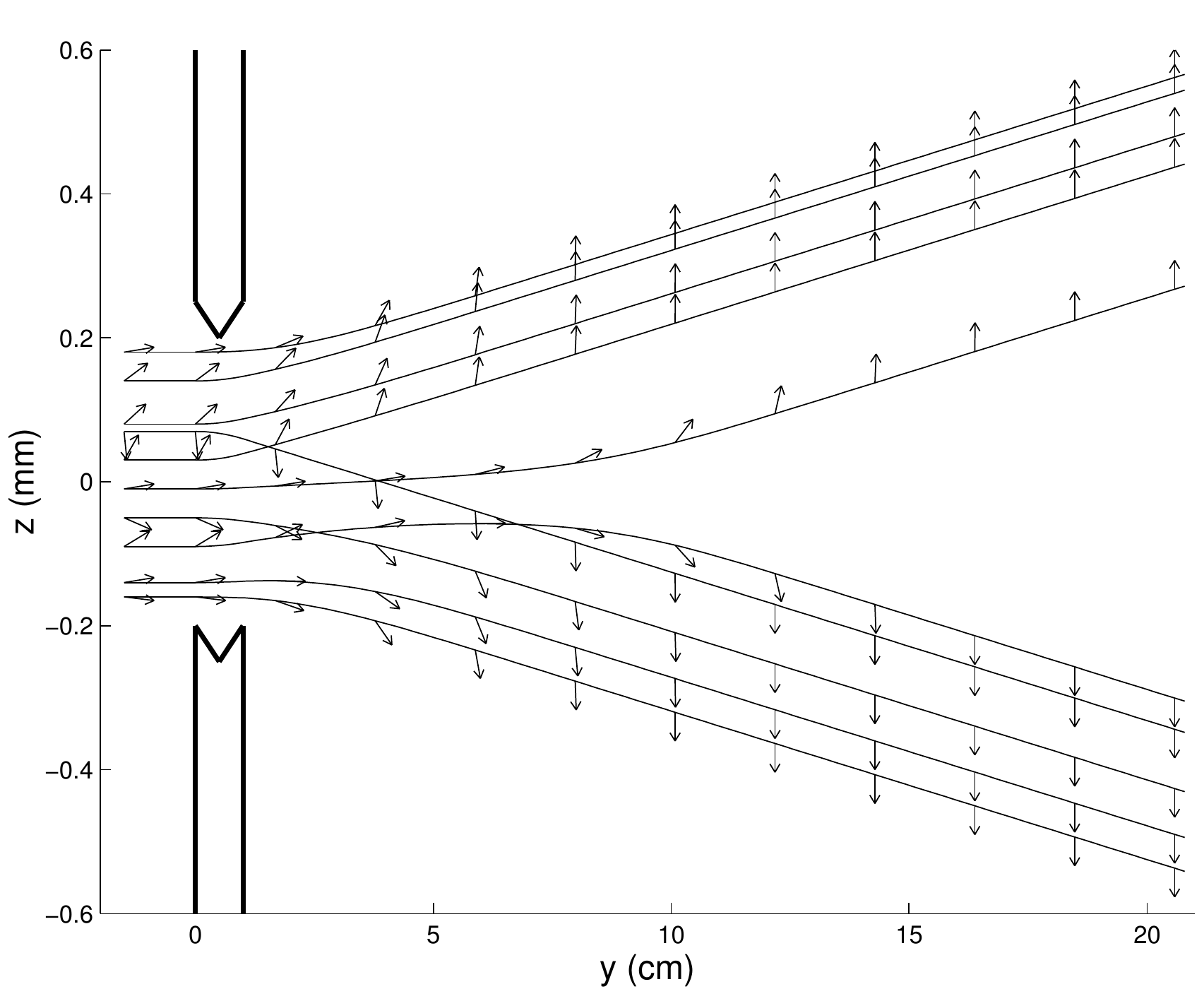}
\caption{\label{fig:SetG-10trajectoires} Ten silver atom
trajectories where the initial orientation ($\theta_0,\varphi_0$)
has been randomly chosen; arrows represent the spin orientation
$\theta(z,t)$ along the trajectories.}
\end{center}
\end{figure}

Now let us consider a mixture of pure states where the initial
orientation ($\theta_0,\varphi_0$) from the spinor has been
randomly chosen. These are the conditions of the initial Stern and
Gerlach experiment. Figure~\ref{fig:SetG-10trajectoires}
represents a simulation of 10 quantum trajectories of silver atoms
from which the initial positions $z_0$ are also randomly chosen.

Finally, the de
Broglie-Bohm trajectories propose a clear interpretation of the spin 
measurement in quantum mechanics. There is interaction with the
measuring apparatus as is generally stated; and there is indeed a
minimum time required to measure. However this measurement and
this time do not have the signification that is usually applied to them.
The result of the Stern-Gerlach experiment is not the measure of
the spin projection along the $z$-axis, but the orientation of the
spin either in the direction of the magnetic field gradient, or in
the opposite direction. It depends on the position of the particle
in the wave function. We have therefore a simple explanation for
the non-compatibility of spin measurements along different axes.
The measurement duration is then the time necessary for the
particle to point its spin in the final direction.

\section{EPR-B experiment}
\label{sect:EPR-B}

Nonseparability is one of the most puzzling aspects of quantum
mechanics. For over thirty years, the EPR-B, the spin version of
the Einstein-Podolsky-Rosen experiment~\cite{EPR} proposed by
Bohm~\cite{Bohm_1951}, the Bell theorem~\cite{Bell64}
and the BCHSH inequalities~\cite{Bell64,BCHSH,Bell_1987} have been
at the heart of the debate on hidden variables and non-locality.
Many experiments since Bell's paper have demonstrated violations
of these inequalities and have vindicated quantum
theory~\cite{Clauser_1972}.
Now, EPR pairs of massive atoms are also
considered~\cite{Beige}. The usual conclusion of these
experiments is to reject the non-local realism for two reasons:
the impossibility of decomposing a pair of entangled atoms into
two states, one for each atom, and the impossibility of
interaction faster than the speed of light.

Here, we show that there exists a de Broglie-Bohm interpretation which answers these two questions positively. To
demonstrate this non-local realism, two methodological conditions
are necessary. The first condition is the same as in the Stern-Gerlach experiment:
the solution
to the entangled state is obtained by resolving the Pauli equation
from an initial singlet wave function with a spatial extension as:
\begin{equation}\label{eq:7psi-0}
    \Psi_{0}(\textbf{r}_A,\textbf{r}_B) =\frac{1}{\sqrt{2}}f(\textbf{r}_A) f(\textbf{r}_B)(|+_{A}\rangle |-_{B}\rangle - |-_{A}\rangle |
    +_{B}\rangle),
\end{equation}
and not from a simplified wave function without spatial extension:
\begin{equation}\label{eq:7psi-1}
    \Psi_{0}(\textbf{r}_A,\textbf{r}_B) =\frac{1}{\sqrt{2}}(|+_{A}\rangle |-_{B}\rangle - |-_{A}\rangle |
    +_{B}\rangle).
\end{equation}
$f$ function and $|\pm\rangle$ vectors are presented later.

The resolution in space of the Pauli equation is essential: it
enables the spin measurement by spatial quantization  and explains
the determinism and the disentangling process. To explain the
interaction and the evolution between the spin of the two
particles, we consider a two-step version of the EPR-B experiment.
It is our second methodological condition. A first causal
interpretation of EPR-B experiment was proposed in 1987 by
Dewdney, Holland and
Kyprianidis~\cite{Dewdney_1987b} using these two conditions. 
However, this interpretation had a flaw~\cite{Holland_1993} (p. 418): the
spin module of each particle depends directly on the singlet wave
function, and thus the spin module of each particle varied during
the experiment from 0 to $\frac{\hbar}{2}$. We present a de 
Broglie-Bohm interpretation that avoid this flaw.~\cite{Gondran_2012}

\begin{figure}[h]
\includegraphics[width=0.95\linewidth]{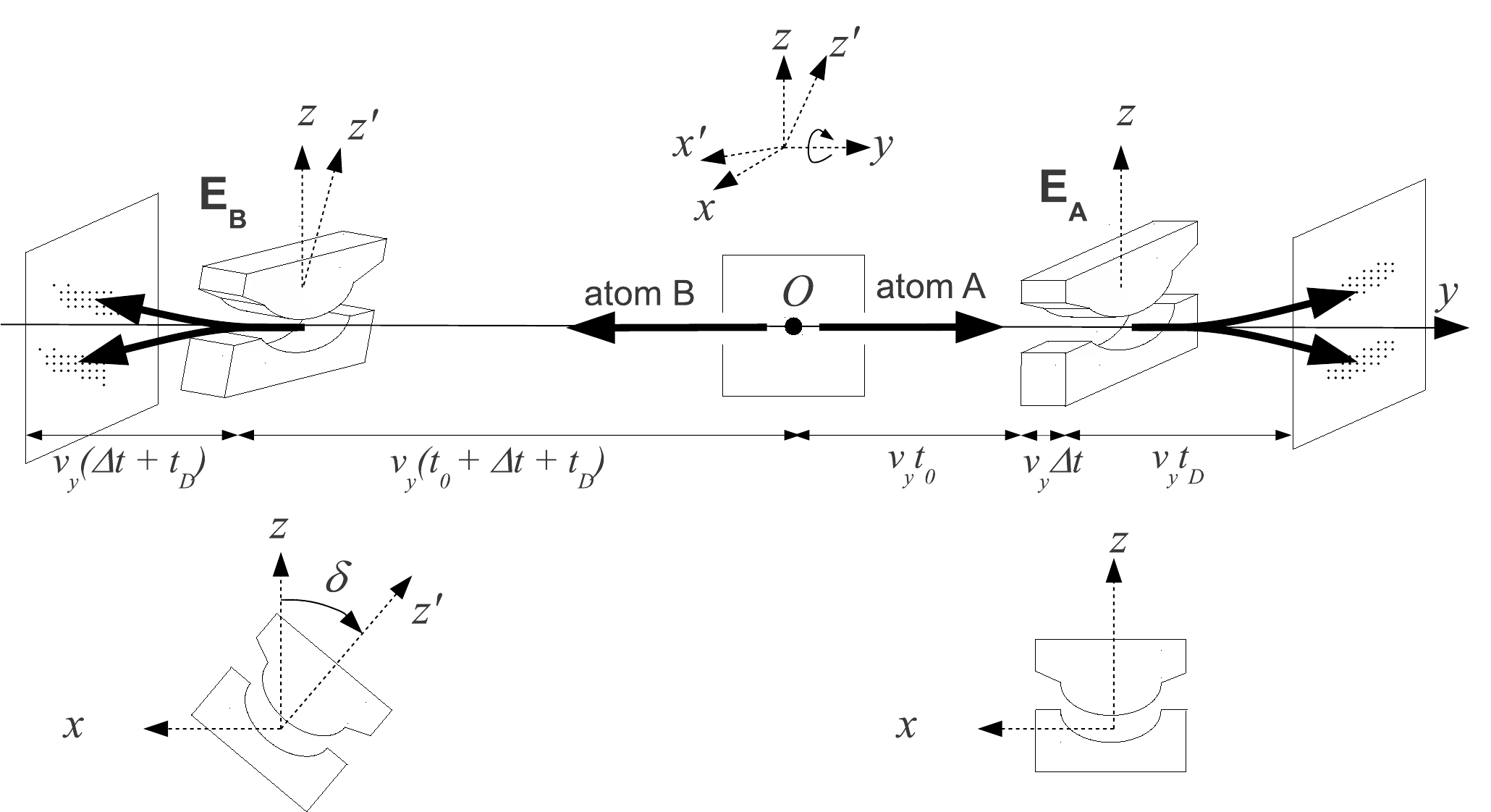}
\caption{\label{fig:expEPR}Schematic configuration of the EPR-B
experiment.}
\end{figure}

Figure \ref{fig:expEPR} presents the Einstein-Podolsky-Rosen-Bohm
experiment. A source $S$ created in O pairs of identical atoms A
and B, but with opposite spins. The atoms A and B split following
the $y$-axis in opposite directions, and head towards two identical
Stern-Gerlach apparatus $\textbf{E}_\textbf{A}$ and $\textbf{E}_\textbf{B}$. The
electromagnet $\textbf{E}_\textbf{A}$ "measures" the spin of A along the
$z$-axis and the electromagnet $\textbf{E}_\textbf{B}$ "measures" the spin of
B along the $z'$-axis, which is obtained after a rotation of an
angle $\delta$ around the $y$-axis. The initial wave function of the
entangled state is the singlet state (\ref{eq:7psi-0}) where
$\textbf{r}=(x,z)$,
$f(\textbf{r})=(2\pi\sigma_{0}^{2})^{-\frac{1}{2}}
 e^{-\frac{x^2 + z^2}{4\sigma_0^2}}$, $|\pm_{A}\rangle$ and $|\pm_{B}\rangle$ are the eigenvectors
of the operators $\sigma_{z_A}$ and $\sigma_{z_B}$: $\sigma_{z_A}
|\pm_{A}\rangle= \pm |\pm_{A}\rangle$, $\sigma_{z_B}
|\pm_{B}\rangle= \pm |\pm_{B}\rangle$. We treat the dependence
with $y$ classically: speed $-v_y$ for A and $v_y$ for B. The wave
function $\Psi(\textbf{r}_A, \textbf{r}_B, t)$ of the two
identical particles A and B, electrically neutral and with
magnetic moments $\mu_0$, subject to magnetic fields
${\textbf{E}_\textbf{A}}$ and ${\textbf{E}_\textbf{B}}$, admits on the basis of
$|\pm_{A}\rangle$ and $|\pm_{B}\rangle$ four components
$\Psi^{a,b}(\textbf{r}_A, \textbf{r}_B, t)$ and satisfies the
two-body Pauli equation~\cite{Holland_1993} (p. 417):
% \begin{widetext}
% \begin{eqnarray}
%     i\hbar \frac{\partial \Psi^{a,b}}{\partial t}
%     =\left(-\frac{\hbar^2}{2 m}\Delta_A -\frac{\hbar^2}{2 m}\Delta_B\right)\Psi^{a,b}
%      +\mu B^{\textbf{E}_\textbf{A}}_j (\sigma_j)_{c}^{a}\Psi^{c,b}
%      +\mu B^{\textbf{E}_\textbf{B}}_j (\sigma_j)_{d}^{b}\Psi^{a,d}\label{eq:7Paulideuxcorps1}
% \end{eqnarray}
% \end{widetext}
\begin{eqnarray}
    i\hbar \frac{\partial \Psi^{a,b}}{\partial t}
    =&&\left(-\frac{\hbar^2}{2 m}\Delta_A -\frac{\hbar^2}{2 m}\Delta_B\right)\Psi^{a,b}\nonumber\\
    &+&\mu B^{\textbf{E}_\textbf{A}}_j (\sigma_j)_{c}^{a}\Psi^{c,b}
     +\mu B^{\textbf{E}_\textbf{B}}_j (\sigma_j)_{d}^{b}\Psi^{a,d}\label{eq:7Paulideuxcorps1}
\end{eqnarray}
with the initial conditions:
\begin{equation}\label{eq:7Paulideuxcorps2}
\Psi^{a,b}(\textbf{r}_A, \textbf{r}_B,
0)=\Psi_0^{a,b}(\textbf{r}_A, \textbf{r}_B)
\end{equation}
where $\Psi_0^{a,b}(\textbf{r}_A, \textbf{r}_B)$ corresponds to
the singlet state (\ref{eq:7psi-0}).

To obtain an explicit solution of the EPR-B experiment, we take
the numerical values of the Stern-Gerlach experiment.

One of the difficulties of the interpretation of the EPR-B
experiment is the existence of two simultaneous measurements. By
doing these measurements one after the other, the interpretation
of the experiment will be facilitated. That is the purpose of the
two-step version of the experiment EPR-B studied below.

\subsection{First step EPR-B: Spin measurement of A}

In the first step we make a Stern and Gerlach "measurement" for
atom A, on a pair of particles A and B in a singlet state. This
is the experiment first proposed in 1987 by Dewdney, Holland and
Kyprianidis.~\cite{Dewdney_1987b}

Consider that at time $t_0$ the particle A arrives at the entrance
of electromagnet $\textbf{E}_\textbf{A}$. After this exit of the magnetic
field $\textbf{E}_\textbf{A}$, at time $t_0+ \triangle t + t$, the wave
function (\ref{eq:7psi-0}) becomes~\cite{Gondran_2012}:
% \begin{equation}\label{eq:7psiexperience1}
% \Psi(\textbf{r}_A, \textbf{r}_B, t_0 + \triangle t+ t )= \frac{1}
% {\sqrt{2}} f(\textbf{r}_B)\times \left( f^{+}(\textbf{r}_A,t)
% |+_{A}\rangle | -_{B}\rangle - f^{-}(\textbf{r}_A,t) |-_{A}\rangle
% | +_{B}\rangle\right)
% \end{equation}
{\small
\begin{eqnarray}
\Psi(\textbf{r}_A, \textbf{r}_B, t_0 + \triangle t+ t )= 
\frac{1}{\sqrt{2}} f(\textbf{r}_B)\times (
&& f^{+}(\textbf{r}_A,t) |+_{A}\rangle | -_{B}\rangle \nonumber\\
&-& f^{-}(\textbf{r}_A,t) |-_{A}\rangle
| +_{B}\rangle)\nonumber\\&&\label{eq:7psiexperience1}
\end{eqnarray}}
with
\begin{equation}\label{eq:7fonction}
f^{\pm}(\textbf{r},t)\simeq f(x, z \mp z_\triangle \mp ut)
e^{i(\frac{\pm muz}{\hbar}+ \varphi^\pm (t))}
\end{equation}
where $ z_{\Delta} $ and $u $ are given by equation (\ref{eq:zdeltavitesse}).

The atomic density $\rho(z_A, z_B,t_0 + \Delta t + t)$ is found by
integrating $\Psi^{*}(\textbf{r}_A,\textbf{r}_B, t_0 + \triangle t
+ t )\Psi(\textbf{r}_A, \textbf{r}_B, t_0 + \triangle t+ t)$ on
$x_A$ and $x_B$:
% \begin{equation} \label{eq:7densitéaprèschampmagnétiqueAB}
%     \rho(z_A, z_B,t_0 + \Delta t+ t) = \left((2\pi\sigma_0^2)^{-\frac{1}{2}}
%                   e^{-\frac{(z_B)^2}{2\sigma_0^2}}\right)
%     \times\left((2\pi\sigma_0^2)^{-\frac{1}{2}}
%                   \frac{1}{2}\left(e^{-\frac{(z_A-z_{\Delta}- ut)^2}{2\sigma_0^2}}+
%                   e^{-\frac{(z_A+z_{\Delta}+
%                   ut)^2}{2\sigma_0^2}}\right)\right).
% \end{equation}
\begin{eqnarray} 
    \rho&(z_A&, z_B,t_0 + \Delta t+ t) = \left((2\pi\sigma_0^2)^{-\frac{1}{2}}
                  e^{-\frac{(z_B)^2}{2\sigma_0^2}}\right)\label{eq:7densitéaprèschampmagnétiqueAB}\\
    &\times&\left((2\pi\sigma_0^2)^{-\frac{1}{2}}
                  \frac{1}{2}\left(e^{-\frac{(z_A-z_{\Delta}- ut)^2}{2\sigma_0^2}}+
                  e^{-\frac{(z_A+z_{\Delta}+
                  ut)^2}{2\sigma_0^2}}\right)\right).\nonumber
\end{eqnarray}
We deduce that the beam of particle A is divided into two, while
the beam of particle B stays undivided:
\begin{itemize}
	\item the density of A is the same, whether particle A is
entangled with B or not, 	\item the density of B is not affected
by the "measurement" of A.
\end{itemize}

Our first conclusion is:  the position of B does not depend
on the measurement of A, only the spins are involved. We conclude
from equation (\ref{eq:7psiexperience1}) that the spins of A and B
remain opposite throughout the experiment. These are the two
properties used in the causal interpretation.

\subsection{Second step EPR-B: Spin measurement of B}

The second step is a continuation of the first and corresponds to
the EPR-B experiment broken down into two steps. On a pair of
particles A and B in a singlet state, first we made a Stern and
Gerlach measurement on the A atom between $t_0$ and $t_0+
\triangle t+ t_D $, secondly, we make a  Stern and Gerlach
measurement on the B atom with an electromagnet $\textbf{E}_\textbf{B}$
forming an angle $\delta$ with $\textbf{E}_\textbf{A}$ during $t_0 +
\triangle t+ t_D$ and $t_0+ 2( \triangle t + t_D)$.

At the exit of magnetic field $\textbf{E}_\textbf{A}$, at time $t_0 +
\triangle t +t_D$, the wave function is given by
(\ref{eq:7psiexperience1}). Immediately after the measurement of
A, still at time $t_0+ \triangle t+ t_D $, the wave function of B
depends on the measurement $\pm $ of A:
\begin{equation}\label{eq:7psiexperience1BcondmesA}
\Psi_{B /\pm A}(\textbf{r}_B, t_0 + \triangle t+ t_1 )=
f(\textbf{r}_B) |\mp_{B}\rangle.
\end{equation}
Then, the measurement of B at time $t_0+ 2( \triangle t + t_D)$
yields, in this two-step version of the EPR-B experiment, the same
results for spatial quantization and correlations of spins as in
the EPR-B experiment.
%%%%%%%%%%%%%%%%%%%%%%%%%%%%%%%%%%%%%%%%%%%%%%%%%%%%%%%%%%%%%%%%%%%%%%%%%%%%%%
%%%%%%%%%%%%%%%%%%%%%%%%%%%%%%%%%%%%%%%%%%%%%%%%%%%%%%%%%%%%%%%%%%%%%%%%%%%%%%
%%%%%%%%%%%%                    Causal interpretation            %%%%%%%%%%%%%
%%%%%%%%%%%%%%%%%%%%%%%%%%%%%%%%%%%%%%%%%%%%%%%%%%%%%%%%%%%%%%%%%%%%%%%%%%%%%%
%%%%%%%%%%%%%%%%%%%%%%%%%%%%%%%%%%%%%%%%%%%%%%%%%%%%%%%%%%%%%%%%%%%%%%%%%%%%%%
\subsection{Causal interpretation of the EPR-B experiment}

We assume, at the creation of the two entangled particles A and B,
that each of the two particles A and B has an initial wave
function with opposite spins: $\Psi_0^A(\textbf{r}_A, \theta^A_0,
\varphi^A_0)= f(\textbf{r}_A)
\left(\cos\frac{\theta^A_0}{2}|+_{A}\rangle +
\sin\frac{\theta^A_0}{2}e^{i \varphi^A_0}|-_{A}\rangle\right)$ and
$\Psi_0^B(\textbf{r}_B , \theta^B_0, \varphi^B_0)= f(\textbf{r}_B)
\left(\cos\frac{\theta^B_0}{2}|+_{B}\rangle +
\sin\frac{\theta^B_0}{2}e^{i \varphi^B_0}|-_{B}\rangle\right)$
with $\theta_0^B= \pi-\theta_0^A$ and $\varphi_0^B= \varphi_0^A
-\pi$. The two particles A and B are statistically prepared as in the Stern and Gerlach experiment. Then the Pauli principle tells us that the two-body wave
function must be antisymmetric; after calculation we find the same
singlet state (\ref{eq:7psi-0}):
% \begin{equation}
%  \Psi_0(\textbf{r}_A,\theta^A, \varphi^A,\textbf{r}_B,\theta^B, \varphi^B)= - e^{i \varphi^A} f(\textbf{r}_A) f(\textbf{r}_B)
%  \times
%  \left(|+_{A}\rangle
% |-_{B}\rangle - |-_{A}\rangle|+_{B}\rangle\right).
% \end{equation}
%{\small
\begin{eqnarray}
 \Psi_0(\textbf{r}_A,\theta^A, \varphi^A,\textbf{r}_B,\theta^B, \varphi^B)=&-& e^{i \varphi^A} f(\textbf{r}_A) f(\textbf{r}_B)\\
 &\times& \left(|+_{A}\rangle
|-_{B}\rangle - |-_{A}\rangle|+_{B}\rangle\right).\nonumber
\end{eqnarray}
Thus, we can consider that the singlet wave function is the wave
function of a family of two fermions A and B with opposite spins:
the direction of initial spin A and B exists, but is not
\textit{known}. It is a local hidden variable which is therefore
necessary to add in the initial conditions of the model.

This is not the interpretation followed by the Bohm 
school~\cite{Dewdney_1987b,Dewdney_1986,Bohm_1993,Holland_1993} in the
interpretation of the singlet wave function; they do not assume the existance of wave functions $\Psi_0^A(\textbf{r}_A, \theta^A_0,
\varphi^A_0) $ and $\Psi_0^B(\textbf{r}_B , \theta^B_0, \varphi^B_0) $ for each particle, but only the singlet state $ \Psi_0(\textbf{r}_A,\theta^A, \varphi^A,\textbf{r}_B,\theta^B, \varphi^B) $. In consequence, they suppose a zero spin for each particle at the initial time and a spin module of each particle varied during
the experiment from 0 to $\frac{\hbar}{2}$~\cite{Holland_1993} (p. 418).

Here, we assume that at the initial time we know the spin of each
particle (given by each initial wave function) and the initial
position of each particle.

\textbf{Step 1: spin measurement of A}

In the equation (\ref{eq:7psiexperience1}) particle A can be
considered independent of B. We can therefore give it the wave
function
% \begin{eqnarray}
% \Psi^A(\textbf{r}_A, t_0+ \triangle t+ t )=
% \cos\frac{\theta_0^A}{2} f^{+}(\textbf{r}_A,t)|+_{A}\rangle +
% \sin\frac{\theta_0^A}{2}e^{i
% \varphi_0^A}f^{-}(\textbf{r}_A,t)|-_{A}\rangle\label{eq:fonctiondondeA}
% \end{eqnarray}
\begin{eqnarray}
\Psi^A(\textbf{r}_A, t_0+ \triangle t+ t )=&&
\cos\frac{\theta_0^A}{2} f^{+}(\textbf{r}_A,t)|+_{A}\rangle\nonumber\\ &&+
\sin\frac{\theta_0^A}{2}e^{i
\varphi_0^A}f^{-}(\textbf{r}_A,t)|-_{A}\rangle\label{eq:fonctiondondeA}
\end{eqnarray}
which is the wave function of a free particle in a Stern Gerlach
apparatus and whose initial spin is given by
($\theta_0^A,\varphi_0^A$). For an initial polarization
($\theta_0^A,\varphi_0^A$) and an initial position ($z_0^A$), we
obtain, in the de Broglie-Bohm interpretation~\cite{Bohm_1993} of the
Stern and Gerlach experiment, an evolution of the position
($z_A(t)$) and of the spin orientation of A
($\theta^A(z_A(t),t)$)~\cite{Gondran_2005b}.

The case of particle B is different. B follows a rectilinear
trajectory with $y_B(t)= v_yt$, $z_B(t)=z_0^B$ and $x_B(t)=x_0^B$.
By contrast, the orientation of its spin moves with the
orientation of the spin of A: $\theta^B(t)= \pi -
\theta^A(z_A(t),t)$ and $\varphi^B(t)= \varphi(z_A(t),t)- \pi$. We
can then associate the wave function:
% \begin{eqnarray}
% \Psi^B(\textbf{r}_B, t_0+ \triangle t+ t )=f(\textbf{r}_B)
% %\nonumber\\
% %&&\times
% \left( \cos\frac{\theta^B(t)}{2} |+_{B}\rangle +
% \sin\frac{\theta^B(t)}{2}e^{i
% \varphi^B(t)}|-_{B}\rangle\right).\label{eq:fonctiondondeB}
% \end{eqnarray}
\begin{eqnarray}
\Psi^B(\textbf{r}_B, t_0+ \triangle t+ t )=f(\textbf{r}_B)
&&\left( \cos\frac{\theta^B(t)}{2} |+_{B}\rangle \right.\label{eq:fonctiondondeB}\\
&&\quad\left.+
\sin\frac{\theta^B(t)}{2}e^{i
\varphi^B(t)}|-_{B}\rangle\right).\nonumber
\end{eqnarray}
This wave function is specific, because it depends upon initial
conditions of A (position and spin). The orientation of spin of
the particle B is driven by the particle A \textit{through the
singlet wave function}. Thus, the singlet wave function is the
non-local variable.

\textbf{Step 2: Spin measurement of B }

At the time $t_0 + \Delta t + t_D$, immediately after the
measurement of A, $\theta^B(t_0 + \Delta t + t_D)= \pi$ or 0 in accordance with the
value of $\theta^A(z_A(t),t)$ and the wave function of B is given
by~(\ref{eq:7psiexperience1BcondmesA}). 
The frame $(Ox'yz')$ corresponds to the frame $(Oxyz)$ 
after a rotation of an angle $\delta$ around the $y$-axis.
$\theta^B$ corresponds to the B-spin angle with the $z$-axis, 
and $\theta'^B$  to the B-spin angle with the $z'$-axis, 
then $\theta'^B(t_0 + \Delta t + t_D)= \pi+\delta$ or $\delta$.
In this second step, we are exactly in the
case of a particle in a simple Stern and Gerlach experiment (with magnet $\textbf{E}_\textbf{B}$)
with a specific initial polarization equal to $\pi+\delta$ or $\delta$ and not random like in step 1. Then, the
measurement of B, at time $t_0+ 2( \triangle t +t_D)$), gives, in
this interpretation of the two-step version of the EPR-B
experiment, the same results as in the EPR-B experiment.

\subsection{Physical explanation of non-local influences}

From the wave function of two entangled particles, we find spins,
trajectories and also a wave function for each of the two
particles. In this interpretation, the quantum particle has a
local position like a classical particle, but it has also a
non-local behavior through the wave function. So, it is the wave
function that creates the non classical properties. We can keep a
view of a local realist world for the particle, but we should add
a non-local vision through the wave function. As we saw in step 1,
the non-local influences in the EPR-B experiment only concern the
spin orientation, not the motion of the particles themselves. 
Indeed only spins are entangled in the wave function~(\ref{eq:7psi-0}) 
not positions and motions like in the initial EPR experiment.
This is a key point in the search for a physical explanation of
non-local influences.

The simplest explanation of this non-local influence is to
reintroduce the concept of ether (or the preferred frame), but a
new format given by Lorentz-Poincaré and by Einstein in
1920\cite{Einstein_1920}: "\textit{Recapitulating, we may say that according to the general theory 
of relativity space is endowed with physical qualities; in this sense, 
therefore, there exists an ether. According to the general theory of 
relativity space without ether is unthinkable; for in such space there 
not only would be no propagation of light, but also no possibility of 
existence for standards of space and time (measuring-rods and clocks), 
nor therefore any space-time intervals in the physical sense. But this 
ether may not be thought of as endowed with the quality characteristic 
of ponderable media, as consisting of parts which may be tracked 
through time. The idea of motion may not be applied to it.}"

Taking into account the new experiments, especially Aspect's
experiments, Popper~\cite{Popper_1982} (p. XVIII) defends a
similar view in 1982:

"\textit{I feel not quite convinced that the experiments are
correctly interpreted; but if they are, we just have to accept
action at a distance. I think (with J.P. Vigier) that this would
of course be very important, but I do not for a moment think that
it would shake, or even touch, realism. Newton and Lorentz were
realists and accepted action at a distance; and Aspect's
experiments would be the first crucial experiment between
Lorentz's and Einstein's interpretation of the Lorentz
transformations.}"

Finally, in the de Broglie-Bohm interpretation, the  EPR-B experiments of non-locality have therefore a
great importance, not to eliminate realism and determinism, but as
Popper said, to rehabilitate the existence of a certain type of
ether, like Lorentz's ether and like Einstein's ether in 1920.

\section{Conclusion}

In the three experiments presented in this article, the variable that is measured in fine is the position of the particle given by this impact on a screen. 
In the double-slit, the set of these positions gives the interferences; 
in the Stern-Gerlach and the EPR-B experiments, it is the position of 
the particle impact that defines the spin value. 

It is this position that the de Broglie-Bohm interpretation 
adds to the wave function to define a complete state of the quantum particle. 
The de Broglie-Bohm interpretation is then based only on the initial 
conditions $\Psi^0(x) $ and $X(0)$ and the evolution 
equations~(\ref{eq:schrodinger1}) and~(\ref{eq:champvitesse}). 
If we add as initial condition the "quantum equilibrium hypothesis"~(\ref{eq:quantumequi}), 
we have seen that we can deduce, for these three examples, 
the three postulates of measurement.
These three postulates are not necessary if we solve the time-dependent Schrödinger equation (double-slit experiment) 
or the Pauli equation with spatial extension (Stern-Gerlach and EPR experiments). However, these simulations enable us to better understand those experiments: 
In the double-slit experiment, the interference phenomena appears only some centimeters after the slits and shows the continuity with classical mechanics;
in the Stern-Gerlach experiment, the spin up/down measurement appears also after a given time, called decoherence time; in the EPR-B experiment, only the spin of B is affected by the spin measurement of A, not its density. 
Moreover, the de Broglie-Bohm 
trajectories propose a clear explanation of the spin measurement in quantum mechanics. 

However, we have seen two very different cases in the measurement process. 
In the first case (double slit experiment), there is no influence of 
the measuring apparatus (the screen) on the quantum particle. In the second 
case (Stern-Gerlach experiment, EPR-B), there is an interaction with the
measuring apparatus (the magnetic field) and the quantum particle. The result 
of  the measurement depends on the position of the particle in the wave function.
The measurement duration is then the time necessary for the stabilisation of the result.

This heterodox interpretation clearly explains experiments with a set of quantum particles that are statistically prepared. These particles verify the "quantum equilibrium hypothesis" and the de Broglie-Bohm interpretation establishes continuity with classical mechanics. However, there is no reason that the de Broglie-Bohm interpretation can be extended to quantum particles that are not statistically prepared. This situation occurs when the wave packet corresponds to a
quasi-classical coherent state, introduced in 1926 by
Schr\"odinger~\cite{Schrodinger_26}.  The field quantum theory and
the second quantification are built on these coherent
states~\cite{Glauber_65}. It is also the case, for the hydrogen atom, of
localized wave packets whose motion are on the classical trajectory
(an old dream of Schr\"odinger's). Their existence was predicted in 1994 by
Bialynicki-Birula, Kalinski, Eberly, Buchleitner and
Delande~\cite{Bialynicki_1994, Delande_1995, Delande_2002}, and
discovered recently by Maeda and Gallagher~\cite{Gallagher} on
Rydberg atoms. For these non statistically prepared quantum particles, we have shown~\cite{Gondran2011,Gondran2012a} that the 
natural interpetation is the Schrödinger interpretation proposed at the Solvay congress in 1927. Everythings happens as if the quantum mechanics interpretation depended on the preparation of the particles (statistically or not statistically prepared). It is perhaps  
a response to the "theory of the double solution" that Louis de Broglie was seeking since 1927: "\textit{I introduced as the "double solution theory" the idea that
it was necessary to distinguish two different solutions that are both linked to the wave equation, one that I called
wave $u$, which was a real physical wave represented by a singularity as it was not normalizable due to a local anomaly defining the particle, the other one as Schr\"odinger's  $\Psi$ wave, which is a probability representation as it is normalizable without singularities.}"~\cite{Broglie}

\appendix

\section{Calculating the spinor evolution in the Stern-Gerlach experiment}

In the magnetic field $B=(B_x,0,B_z)$, the Pauli equation
(\ref{eq:Pauli}) gives coupled Schrödinger equations for each
spinor component
% \begin{widetext}
% \begin{equation}\label{eq:Paulicomplet}
%      i\hbar \frac{\partial\psi_{\pm}}{\partial t}(x,z,t)= - \frac{\hbar^2}{2 m} \nabla^{2} \psi_{\pm }(x,z,t)  \pm
%      \mu_B (B_0 - B_0' z) \psi_{\pm}(x,z,t) \mp i \mu_B B_0' x
%      \psi_{\mp}(x,z,t).
% \end{equation}
% \end{widetext}
\begin{eqnarray}\label{eq:Paulicomplet}
    i\hbar \frac{\partial\psi_{\pm}}{\partial t}(x,z,t)= 
    &-& \frac{\hbar^2}{2 m} \nabla^{2} \psi_{\pm }(x,z,t)  \nonumber\\
    &\pm&\mu_B (B_0 - B_0' z) \psi_{\pm}(x,z,t) \nonumber\\
    &\mp&i \mu_B B_0' x\psi_{\mp}(x,z,t).
\end{eqnarray}

If one effects the transformation \cite{Platt_1992}
\begin{equation}\nonumber
     \psi_{\pm}(x,z,t)= \exp \left(\pm \frac{i \mu_B B_0 t}{\hbar}\right)
     \overline{\psi}_{\pm}(x,z,t)
\end{equation}
equation~(\ref{eq:Paulicomplet}) becomes
% \begin{widetext}
% \begin{equation}\nonumber
%      i\hbar \frac{\partial\overline{\psi}_{\pm}}{\partial t}(x,z,t)= -\frac{\hbar^{2}}{2 m} \nabla^{2} \overline{\psi}_{\pm}(x,z,t)
%      \mp \mu_B
%      B'_0 z \overline{\psi}_{\pm}(x,z,t) \mp i \mu_B
%      B'_0 x \overline{\psi}_{\mp}(x,z,t)\exp(\pm i \frac{2 \mu_B B_0 t}{\hbar})
% \end{equation}
% \end{widetext}
\begin{eqnarray*}
     i\hbar \frac{\partial\overline{\psi}_{\pm}}{\partial t}(x,z,t)= 
     &-&\frac{\hbar^{2}}{2 m} \nabla^{2} \overline{\psi}_{\pm}(x,z,t)\\
     &\mp& \mu_B B'_0 z \overline{\psi}_{\pm}(x,z,t) \\
     &\mp& i \mu_B B'_0 x \overline{\psi}_{\mp}(x,z,t)\exp\left(\pm i \frac{2 \mu_B B_0 t}{\hbar}\right)
\end{eqnarray*}
The coupling term oscillates rapidly with the Larmor frequency
$\omega_{L}= \frac{2 \mu_B B_0}{\hbar}=1,4 \times 10^{11} s^{-1}$.
Since $|B_{0}|\gg |B'_{0}z|$ and $|B_{0}|\gg |B'_{0}x|$, the
period of oscillation is short compared to the motion of the wave
function. Averaging over a period that is long compared to the
oscillation period, the coupling term vanishes, which
entails\cite{Platt_1992}
\begin{equation}\label{eq:Paulimoyen}
     i\hbar \frac{\partial\overline{\psi}_{\pm}}{\partial t}(x,z,t)= -\frac{\hbar^{2}}{2 m} \nabla^{2} \overline{\psi}_{\pm}(x,z,t)
     \mp \mu_B
     B'_0 z \overline{\psi}_{\pm}(x,z,t).
\end{equation}

Since the variable x is not involved in this equation and
$\psi^{0}_{\pm}(x,z)$ does not depend on x,
$\overline{\psi_{\pm}}(x,z,t) $ does not depend on x:
$\overline{\psi_{\pm}}(x,z,t)\equiv\overline{\psi_{\pm}}(z,t) $.
Then we can explicitly compute the preceding equations for all t
in $[0, \Delta t]$ with $\Delta t=\frac{\Delta l}{v}=2 \times
10^{5}s$.

We obtain:
% \begin{equation}\nonumber
% \overline{\psi}_{+}(z,t)= \psi_{K}(z,t) \cos \frac{\theta_0}{2}e^{
% i\frac{\varphi_0}{2}}~~~~~\text{and}~~K=-\mu_B B'_{0}
% \end{equation}
% 
% \begin{equation}\nonumber
% \overline{\psi}_{-}(z,t)= \psi_{K}(z,t)
% i\sin\frac{\theta_0}{2}e^{-i\frac{\varphi_0}{2}}~~~~~\text{and}~~K=+\mu_B
% B'_{0}
% \end{equation}
\begin{eqnarray*}
&\overline{\psi}_{+}(z,t)= \psi_{K}(z,t) \cos \frac{\theta_0}{2}e^{i\frac{\varphi_0}{2}}~~~&~~\text{and}~~K=-\mu_B B'_{0}\\
&\overline{\psi}_{-}(z,t)= \psi_{K}(z,t) i\sin\frac{\theta_0}{2}e^{-i\frac{\varphi_0}{2}}&~~\text{and}~~K=+\mu_B B'_{0}
\end{eqnarray*}
$\sigma_t^2 = \sigma_0^2 + \left(\frac{\hbar
t}{2m\sigma_0}\right)^2$ and
% \begin{equation}\label{eq:ondegauss}
%  \psi_{K}(z,t)=(2\pi\sigma_{t}^{2})^{-\frac{1}{4}}
%                       e^{-\frac{(z+\frac{K t^{2}}{2 m})^2}{4\sigma_t^2}}\exp \frac{i}{\hbar}
%                       \left[-\frac{\hbar}{2}\tan^{-1}(\frac{\hbar t}{2 m \sigma_{0}^{2}})
%                       -Ktz-\frac{K^{2}t^{3}}{6 m}
%                       +\frac{(z+\frac{K t^{2}}{2 m})^{2}\hbar^{2}t^{2}}{8 m \sigma_{0}^{2}
%                       \sigma_{t}^{2}}\right].
% \end{equation}
{\small
\begin{eqnarray}
 \psi_{K}(z,t)&=&(2\pi\sigma_{t}^{2})^{-\frac{1}{4}}
                      e^{-\frac{(z+\frac{K t^{2}}{2 m})^2}{4\sigma_t^2}}
\exp \frac{i}{\hbar}
                      \left[-\frac{\hbar}{2}\tan^{-1}\left(\frac{\hbar t}{2 m \sigma_{0}^{2}}\right)\right.\nonumber\\
                      &&\left.-Ktz-\frac{K^{2}t^{3}}{6 m}
                      +\frac{(z+\frac{K t^{2}}{2 m})^{2}\hbar^{2}t^{2}}{8 m \sigma_{0}^{2}
                      \sigma_{t}^{2}}\right].\label{eq:ondegauss}
\end{eqnarray}}
where~(\ref{eq:ondegauss}) is a classical result.\cite{Feynman}

The experimental conditions give $\frac{\hbar \Delta t}{2 m
\sigma_0}=4 \times 10^{-11}~m \ll \sigma_{0}=10^{-4}~m$. We deduce
the approximations $\sigma_{t} \simeq  \sigma_0$ and
\begin{equation}\label{eq:ondegaussapprox}
 \overline{\psi}_{K}(z,t)  \simeq (2\pi\sigma_{0}^{2})^{-\frac{1}{4}}
                      e^{-\frac{(z+\frac{K t^{2}}{2 m})^2}{4\sigma_0^2}}\exp \frac{i}{\hbar}
                      \left[ -Ktz-\frac{K^{2}t^{3}}{6 m}\right].
\end{equation}

At the end of the magnetic field, at time $\Delta t$, the spinor
equals to
\begin{equation}\label{eq:ondeadeltat}
    \Psi (z,\Delta t) = \left( \begin{array}{c}
                                \psi_{+}(z,\Delta t)\\
                                \psi_{-}(z,\Delta t)
                            \end{array}
                     \right)
\end{equation}
with
\begin{eqnarray*}
\psi_{+}(z,\Delta t)&=&(2\pi\sigma_{0}^{2})^{-\frac{1}{4}}
                      e^{-\frac{(z - z_{\Delta})^2}{4\sigma_0^2}+ \frac{i}{\hbar}m u z}  \cos \frac{\theta_0}{2}e^{i \varphi_{+}}\\
\psi_{-}(z,\Delta t)&=& (2\pi\sigma_{0}^{2})^{-\frac{1}{4}}
                      e^{-\frac{(z + z_{\Delta})^2}{4\sigma_0^2}- \frac{i}{\hbar}m u z} i\sin\frac{\theta_0}{2}e^{i \varphi_{-}}
\end{eqnarray*}

% \begin{equation}\nonumber
%     z_{\Delta}=\frac{\mu_B B'_{0}(\Delta
%     t)^{2}}{2 m},~~~~~~u =\frac{\mu_0 B'_{0}(\Delta t)}{m}~~~~and
% \end{equation}
% \begin{equation}\nonumber
%     \varphi_{+}=\frac{\varphi_{0}}{2} -\frac{\mu_{B}B_0 \Delta
%     t}{\hbar}-\frac{K^{2}(\Delta t)^{3}}{6 m \hbar};~~~~\varphi_{-}=-\frac{\varphi_{0}}{2} +\frac{\mu_{0}B_0 \Delta
%     t}{\hbar}-\frac{K^{2}(\Delta t)^{3}}{6 m \hbar}.
% \end{equation}

\begin{eqnarray*}
    &z_{\Delta}&=\frac{\mu_B B'_{0}(\Delta
    t)^{2}}{2 m},~~~~~~u =\frac{\mu_0 B'_{0}(\Delta t)}{m}~~~~\text{and}\\
    &\varphi_{+}&=\frac{\varphi_{0}}{2} -\frac{\mu_{B}B_0 \Delta
    t}{\hbar}-\frac{K^{2}(\Delta t)^{3}}{6 m \hbar};\\
    &\varphi_{-}&=-\frac{\varphi_{0}}{2} +\frac{\mu_{0}B_0 \Delta
    t}{\hbar}-\frac{K^{2}(\Delta t)^{3}}{6 m \hbar}.
\end{eqnarray*}

We remark that the passage through the magnetic field gives the
equivalent of a velocity $+u$ in the direction $0z$ to the
function $\psi_+$ and a velocity $-u$ to the function $\psi_{-}$.
Then we have a free particle with the initial wave
function~(\ref{eq:ondeadeltat}). The Pauli equation resolution
again yields $\psi_{\pm}(x,z,t)=\psi_{x}(x,t) \psi_{\pm}(z,t)$ and
with the experimental conditions we have $\psi_{x}(x,t) \simeq
(2\pi\sigma_{0}^{2})^{-\frac{1}{4}} e^{-\frac{x^2}{4\sigma_0^2}}$
and
% \begin{equation}\nonumber
% \psi_{+}(z,t +\Delta t) \simeq (2\pi\sigma_{0}^{2})^{-\frac{1}{4}}
%                       e^{-\frac{(z - z_{\Delta}- u t)^2}{4\sigma_0^2}+ \frac{i}{\hbar}(m u z -\frac{1}{2}m u^2 t + \hbar \varphi_{+})}
%                         \cos \frac{\theta_0}{2}
% \end{equation}
% \begin{equation}\nonumber
% \psi_{-}(z,t + \Delta t) \simeq
% (2\pi\sigma_{0}^{2})^{-\frac{1}{4}}
%                       e^{-\frac{(z + z_{\Delta}+u t)^2 }{4\sigma_0^2}+\frac{i}{\hbar}(- m u z -\frac{1}{2}m u^2 t + \hbar \varphi_{-})}
%                        i\sin\frac{\theta_0}{2}
% \end{equation}
\begin{eqnarray*}
\psi_{+}(z,t +\Delta t)  &\simeq&(2\pi\sigma_{0}^{2})^{-\frac{1}{4}}\cos \frac{\theta_0}{2}\\
                         &&\times \exp^{-\frac{(z - z_{\Delta}- u t)^2}{4\sigma_0^2}+ \frac{i}{\hbar}(m u z -\frac{1}{2}m u^2 t + \hbar \varphi_{+})}\\                        
\psi_{-}(z,t + \Delta t) &\simeq&(2\pi\sigma_{0}^{2})^{-\frac{1}{4}}i\sin\frac{\theta_0}{2}\\
                         &&\times \exp^{-\frac{(z + z_{\Delta}+u t)^2 }{4\sigma_0^2}+\frac{i}{\hbar}(- m u z -\frac{1}{2}m u^2 t + \hbar \varphi_{-})}
\end{eqnarray*}

\end{document}